\documentclass[fleqn,usenatbib]{mnras}

\usepackage{newtxtext,newtxmath}

\usepackage[T1]{fontenc}

\DeclareRobustCommand{\VAN}[3]{#2}
\let\VANthebibliography\thebibliography
\def\thebibliography{\DeclareRobustCommand{\VAN}[3]{##3}\VANthebibliography}

\usepackage{graphicx}	
\usepackage{amsmath}	
\usepackage{amssymb}	

\newcommand{\Gyr}{\,{\rm Gyr}}
\newcommand{\Myr}{\,{\rm Myr}}

\newcommand{\hii}{\mbox{H\,{\sc ii}}}

\def \asec{$^{\prime\prime}$}

\newcommand{\hbeta}{\mbox{H\,{\sc $\beta$}}}
\newcommand{\halpha}{\mbox{H\,{\sc $\alpha$}}}
\newcommand{\oiii}{\mbox{[O\,{\sc iii]}{$\lambda\lambda 4959,5007$}}}
\newcommand{\oiiia}{\mbox{[O\,{\sc iii]}{$\lambda 5007$}}}
\newcommand{\oii}{\mbox{[O\,{\sc ii]}{$\lambda\lambda 3726,3729$}}}

\newcommand{\nii}{\mbox{[N\,{\sc ii]}{$\lambda 6584$}}}
\newcommand{\neiii}{\mbox{[Ne\,{\sc iii]}{$\lambda 3870$}}}
\newcommand{\sii}{\mbox{[S\,{\sc ii]}{$\lambda\lambda 6717,6731$}}}
\newcommand{\ciii}{\mbox{C\,{\sc iii]}{$\lambda 1907,1909$}}}

\newcommand{\civ}{\mbox{C\,{\sc iv}{$\lambda\lambda 1548,1550$}}}
\newcommand{\heii}{\mbox{He\,{\sc ii}{$\lambda 1640$}}}
\newcommand{\nv}{\mbox{N\,{\sc v}{$\lambda\lambda 1238,1242$}}}

\newcommand{\oiiinwl}{\mbox{[O\,{\sc iii]}}}
\newcommand{\oiinwl}{\mbox{[O\,{\sc ii]}}}
\newcommand{\neiiinwl}{\mbox{[Ne\,{\sc iii]}}}

\newcommand{\heiinwl}{\mbox{He\,{\sc ii}}}

\newcommand{\zstar}{$Z_{\star}$}
\newcommand{\zgas}{$Z_{\mathrm{g}}$}

\newcommand{\mstar}{$M_{\star}$}
\newcommand{\msun}{$\mathrm{M}_{\odot}$}

\newcommand{\mzrstar}{MZR$_{\star}$}
\newcommand{\mzrgas}{MZR$_{\mathrm{g}}$}

\newcommand{\oferel}{$\mathrm{[Fe/H]}-\mathrm{[O/Fe]}$}

\title[$\alpha$-enhancement in star-forming galaxies at $z\simeq3.4$]{{The NIRVANDELS Survey: a robust detection of $\mathbf{\alpha}$-enhancement in star-forming galaxies at $\mathbf{z\simeq3.4}$}}

\author[F. Cullen et al.]{F. Cullen$^{1}$\thanks{E-mail: fc@roe.ac.uk},
A. E. Shapley$^{2}$,
R. J. McLure$^{1}$,
J. S. Dunlop$^{1}$,
R. L. Sanders$^{3}$,
M. W. Topping$^{2}$, 
N. A. Reddy$^{4}$,\and
R. Amor\'in${^{5,6}}$,
R. Begley$^{1}$,
M. Bolzonella${^{7}}$,
A. Calabr\`o${^{8}}$,
A. C. Carnall${^{1}}$,
M. Castellano${^{8}}$, 
A. Cimatti${^{9,10}}$, \and
M. Cirasuolo$^{11}$,
G. Cresci${^{10}}$,
A. Fontana${^{8}}$,
F. Fontanot${^{13}}$,
B. Garilli${^{12}}$,
L. Guaita${^{14}}$, 
M. Hamadouche$^1$,\and
N. P. Hathi${^{15}}$, 
F. Mannucci${^{10}}$, 
D. J. McLeod${^{1}}$,
L. Pentericci$^{8}$,
A. Saxena${^{16}}$,
M. Talia${^{7,9}}$,
G. Zamorani${^{7}}$
\\
Affiliations are listed at the end of the paper}

\date{Accepted ---. Received ---; in original form ---}

\pubyear{2020}

\begin{document}
\label{firstpage}
\pagerange{\pageref{firstpage}--\pageref{lastpage}}
\maketitle

\begin{abstract}
We present results from the NIRVANDELS survey on the gas-phase metallicity (\zgas, tracing O/H) and stellar metallicity (\zstar, tracing Fe/H) of 33 star-forming galaxies at redshifts $2.95 < z < 3.80$.
Based on a combined analysis of deep optical and near-IR spectra, tracing the rest-frame far ultraviolet (FUV; $1200-2000${\AA}) and rest-frame optical ($3400-5500${\AA}) respectively, we present the first simultaneous determination of the stellar and gas-phase mass-metallicity relationships (MZRs) at $z\simeq3.4$.
In both cases, we find that metallicity increases with increasing stellar mass (\mstar), and that the power-law slope at \mstar$\lesssim 10^{10} \mathrm{M}_{\odot}$ of both MZRs scales as $Z \propto M_{\star}^{0.3}$.
Comparing the stellar and gas-phase MZRs, we present direct evidence for super-solar O/Fe ratios (i.e., $\alpha$-enhancement) at $z>3$, finding $\mathrm{(O/Fe)} = 2.54 \pm 0.38 \times \mathrm{(O/Fe)}_{\odot}$, with no clear dependence on \mstar.
\end{abstract}




\section{Introduction}\label{sec:intro}

The metal content of galaxies is affected by past and current star formation, gas accretion and galactic winds, and therefore constrains all aspects of the cosmic baryon cycle.
Of particular interest is the evolution of galaxy metallicity with cosmic time, and the scaling relations between metallicity and other galaxy properties, most notably stellar mass (\mstar). 
Accurately determined, these relations and their redshift evolution serve as powerful tests of theoretical models of galaxy evolution \citep{maiolino2019}.

Current constraints on galaxy metallicities primarily come from observations of strong nebular emission lines emitted at rest-frame optical wavelengths ($\simeq3500-7000${\AA}) .
When measured from integrated spectra of star-forming galaxies, the ratios of these emission lines are sensitive to the gas-phase oxygen abundance (O/H) of galactic \hii \ regions \citep{kewley2019}.
Using this technique, gas-phase metallicities (\zgas) have been measured for sizable samples of galaxies from the local Universe out to $z\simeq4$ \citep[e.g.][]{tremonti2004,sanders2020}.
Based on these observations, it is now well-established that \zgas \ is primarily correlated with \mstar \ at all redshifts (i.e., the mass-metallicity relationship, or MZR) with strong evidence for a secondary dependence on star-formation rate (SFR) or molecular gas fraction \citep[i.e., the `fundamental metallicity relationship', or FMR;][]{mannucci2010}.

An alternative approach to measuring galaxy metallicities at $z>2$ utilizes observations of the stellar continuum at far-ultraviolet wavelengths (FUV; $1000-2000${\AA}).
In contrast to nebular emission-line measurements, these estimates of the stellar metallicity (\zstar) are sensitive to the photospheric iron abundance (Fe/H) of O- and B-type stars within galaxies \citep{leitherer2010}.
Early efforts to determine stellar metallicities from FUV spectra at high redshift were limited by small samples sizes \citep[e.g.,][]{halliday2008,sommariva2012}.
However, in \citet{cullen2019}, building upon this early work, we utilized the large number of ultra-deep rest-frame FUV spectra provided by the VANDELS survey \citep{mclure_vandels} to publish the first determination of the stellar mass-metallicity relationship at $z>2$.
We found that the stellar mass-metallicity relationship at $z\simeq3.5$ has a similar shape but lower overall normalization when compared to the local relation, mirroring the redshift evolution observed in the gas-phase relation \citep[see also][]{calabro2021}.

Access to deep rest-frame FUV and optical spectra at $z > 2$ (from ground and space-based optical and near-IR spectroscopy respectively), offers a unique opportunity to move beyond single-element abundances, enabling the study of abundance ratios at early cosmic epochs.
Specifically, the combined analysis of FUV+optical spectra allows for the simultaneous determination of \zgas \ and \zstar, tracing the ratio of oxygen to iron (O/Fe) in young stars and the surrounding ISM \citep{steidel2016}.
The O/Fe ratio is of interest because it is a sensitive probe of the star-formation and chemical enrichment history of galaxies.
Typical O/Fe ratios at high redshift are expected to be enhanced relative to the solar value due to the fact that the element abundance ratios in relatively young star-forming systems will be dominated by core-collapse supernova (CCSNe) yields \citep{maiolino2019}.
Indeed, this result has already been reported in the literature. 
Based on a simultaneous rest-frame FUV + optical analysis of $30$ star-forming galaxies at $z\simeq2.4$, \citet{steidel2016} found $\mathrm{O/Fe}\approx 4-5 \times \mathrm{O/Fe}_{\odot}$.
Similar levels of O/Fe-enhancement have been reported for individual galaxies at $z\simeq2.3$ in more recent studies \citep{topping2020a,topping2020b}.

The observed O/Fe enhancement (also referred to as $\alpha$-enhancement) has a number of important implications.
Firstly, it places robust constraints on the typical star-formation histories at early cosmic epochs, confirming results from previous photometric analyses that the stellar populations at high redshift are typically $< 1$ \Gyr \ old \citep{reddy2012}.
Secondly, a deficit of Fe relative to O will mean that the stellar ionizing spectrum of typical high-redshift galaxies is harder at fixed oxygen abundance compared to galaxies in the local Universe.
This relative hardening of the ionizing continuum at fixed O/H offers a natural explanation for the observed offset of $z>2$ star-forming galaxies relative to $z\simeq0$ galaxies in the common line ratio diagrams \citep[e.g., the BPT diagram;][]{topping2020a, runco2021}.
Non-solar $\alpha$/Fe ratios will also force us to re-think current stellar population techniques when applied to high-redshift galaxies. 
Almost universally, current stellar population synthesis models assume solar abundance ratios; accurate analyses of FUV-optical spectra at $z>6$ with \emph{JWST} will require new models allowing for non-solar abundance ratios.

In this paper, we expand upon a number of previous works at $z\simeq2.5$ \citep{steidel2016,topping2020a,topping2020b} and present a simultaneous analysis of FUV and optical spectra for a sample of $33$ star-forming galaxies drawn from the VANDELS survey at $z\simeq3.4$.
Combining ultra-deep optical VIMOS/VANDELS spectroscopic observations (tracing the rest-frame FUV) with  MOSFIRE $H-$ and $K-$band near-IR follow-up (tracing the rest-frame optical), our analysis provides the first investigation of \zgas \ and \zstar \ for galaxies at $z>3$, and we present the first estimates of O/Fe at these redshifts.
In addition, we present both the stellar and gas-phase MZRs for our sample, tracing O/Fe as a function of stellar mass.

The structure of this paper is as follows.
In Section \ref{sec:data} we discuss our combined VANDELS+MOSFIRE spectroscopic dataset and describe our final $z\simeq3.4$ galaxy sample (referred to throughout the rest of this paper as the NIRVANDELS sample).
In Section \ref{sec:metallicity_derivation} we outline the methods used to determine \zstar \ and \zgas \ from the rest-frame FUV and optical spectroscopy respectively.
In Section \ref{sec:results} we present our determination of the stellar and gas-phase MZRs at $z>3$ along with an estimate of the typical O/Fe ratios of our sample.
In Section \ref{sec:discussion} we discuss some of the implications of our results before summarizing our main conclusions in Section \ref{sec:conclusions}.
Throughout this paper, metallicities are quoted relative to the solar abundance taken from \citet{asplund2009} which has a bulk composition by mass of $Z_{\ast}=0.0142$. 
We assume the following cosmology: $\Omega_{M} =0.3$, $\Omega_\Lambda =0.7$, $H_0 =70$ km s$^{-1}$ Mpc$^{-1}$.

\section{Data and Sample Properties}\label{sec:data}

\subsection{Rest-frame UV VANDELS sample and observations}
The star-forming galaxy sample presented in this paper was initially drawn from the VANDELS ESO public spectroscopic survey \citep{pentericci_vandels,mclure_vandels}.
VANDELS is an ultra-deep, optical, spectroscopic survey of the CANDELS CDFS and UDS fields \citep{grogin2011,koekemoer2011} undertaken using the VIMOS spectrograph on ESO's Very Large Telescope (VLT).
The three categories of VANDELS targets were massive passive galaxies at $1.0 \leq z \leq 2.5$, bright star-forming galaxies at $2.4 \leq z \leq 5.5$ and fainter star-forming galaxies at $3.0 \leq z \leq 7.0$, with the main focus being star-forming galaxies at $z>2.4$ ($85\%$ of targets).
Observations were obtained using the VIMOS medium-resolution grism covering the wavelength range $4800\mbox{\AA} < \lambda_{\mathrm{obs}} < 10000$ {\AA} at a resolution of $R=580$ (with 1.0\asec slits) and a dispersion of 2.5~\AA\ per pixel.

At the redshifts of interest for our study ($2.95 \leq z \leq 3.80$) the VIMOS spectra cover rest-frame UV emission at $\simeq 1000-2000${\AA}, a wavelength range sensitive to various continuum and emission-line features that trace the properties of young, massive stellar populations in star-forming galaxies \citep{cullen2019,cullen2020,calabro2021}.
The observations and reduction of the VIMOS spectra are described in detail in the VANDELS data release papers \citep{pentericci_vandels,garilli2021}. 
The selection of VANDELS targets for near-IR follow-up with MOSFIRE is described below.

\subsection{Rest-frame Optical MOSFIRE sample and observations}
In order to characterize simultaneously the properties of massive stars and ionized gas in $z\simeq 3.4$ star-forming galaxies, we selected a sample of galaxies from the VANDELS survey for near-infrared spectroscopic follow-up observations with the Multi-object Spectrometer for Infrared Exploration \citep[MOSFIRE;][]{mclean2012} on the Keck~I telescope.
The requirement for strong rest-frame optical features to fall within windows of atmospheric transmission translates into discrete allowed redshift ranges for targets for ground-based near-infrared spectroscopic follow-up, including $2.95 \leq z \leq 3.80$ for $z\sim 3$ targets and $2.09 \leq z \leq 2.61$ for those at $z\sim 2$.

For most of our MOSFIRE mask design (i.e., for masks observed in November 2019), we prioritized VANDELS galaxies with $H_{\rm AB}\leq 25.5$, and robustly measured redshifts \citep[e.g., characterized by redshift flags 3, 4, 9, and 14, as defined in][]{pentericci_vandels} at $2.95 \leq z \leq 3.80$.
Slightly higher priority was given to sources with measured \ciii \ rest-UV nebular emission. 
VANDELS sources with robustly measured redshifts (same redshift flags as above) at $2.09 \leq z \leq 2.61$ were targeted with lower priority, and, finally, VANDELS sources with $H_{\rm AB}> 25.5$ and either $2.95 \leq z \leq 3.80$ or $2.09 \leq z \leq 2.61$ were considered the lowest priority.
A slightly different priority scheme was used for the gs\_al1 mask, which was designed a year earlier, prior to the completion of the final VANDELS redshift measurements.
Accordingly, for this mask, galaxies at $H_{\rm AB}\leq 25.5$ and robustly measured redshifts (again, as defined above) at either $2.95 \leq z \leq 3.80$ or $2.09 \leq z \leq 2.61$ were prioritized, followed by VANDELS targets with redshifts yet to be measured. In attempting to optimize the number of VANDELS sources per MOSFIRE pointing that satisfied the above criteria, we found that the best mask configurations contained 15--20 such targets. 

We obtained Keck/MOSFIRE  $H$-, and $K$-band rest-frame optical spectra for the selected VANDELS targets in the GOODS-S and UDS fields. 
At $2.95\leq z\leq3.8$ \hbeta \ and \oiii \ fall in the $K$ band, while \oii \ and \neiii \ fall in the $H$ band. At $2.09 \leq z \leq 2.61$, \halpha, \nii, and \sii \ fall in the $K$ band, while \hbeta \ and \oiii \ fall in the $H$ band.
We collected observations of three MOSFIRE slitmasks (gs\_al1, ud\_van7, and gs\_van2) on  21 October 2018, 13 January 2019, 4 November 2019, and 13 November 2019. Conditions were clear during the observations, with seeing ranging from 0.5\asec to 0.6\asec.
The slitwidth was 0.7\asec, yielding a spectral resolution of $\simeq 3650$ in $H$ and $\simeq 3600$ in $K$. \footnote{While the MOSFIRE slitwidth (0.7\asec) is narrower than the VIMOS slitwidth (1.0\asec), the typical seeing in our MOSFIRE observations (0.5\asec-0.6\asec) was also better than the typical seeing for the VANDELS VIMOS observations (0.7\asec). In fact, the VIMOS and MOSFIRE spectroscopic observations presented here probe comparable intrinsic regions of our target galaxies.} A total of 50 VANDELS sources were targeted within the three masks.
The MOSFIRE observations are summarized in Table \ref{tab:observations}.

We reduced the raw data to produce two-dimensional science and error spectra using the pipeline described in \citet{kriek2015}.
We then optimally extracted one-dimensional science and error spectra from the two-dimensional spectra.
Flux calibrations and slit-loss corrections for each filter were applied as described in \citet{kriek2015} and \citet{reddy2015}. 
Of the 50 VANDELS sources targeted, we obtained rest-frame optical spectra that yielded robust, science-grade redshifts for 35 sources at $2.95\leq z\leq3.8$ and 10 sources at $2.09 \leq z \leq 2.61$.
In this paper, we focus exclusively on the $z\geq2.95$ sources. 


\begin{table*}
\caption{Summary of MOSFIRE Observations}
\label{tab:observations}
\begin{tabular}{cccrrrccccr}
\hline
Mask & R.A.    & Decl.   & P.A.  & $K$ Exptime & $H$ Exptime & $N_{{\rm targ}}$ & $N_{z\sim3}$ & $N_{z\sim2}$ & Obs. Run & Seeing\\
     & (J2000) & (J2000) & (deg) & (sec)       & (sec)        &                  &              &              &          &       \\
\hline
\hline
gs\_al1  & 03:32:43.00 & $-$27:46:25.8 & 261.7 & 11520 & 7080 & 10 & 6 & 2 & 2018 Oct, 2019 Jan& 0.6\asec\\
gs\_van2 & 03:32:10.99 & $-$27:44:11.7 & 8.9   & 7200  & 7200 & 23 & 15 & 6 & 2019 Nov & 0.5\asec\\
ud\_van7 & 02:18:00.43 & $-$05:09:59.1 & 90.1  & 8640  & 8640 & 17 & 14 & 2 & 2019 Nov & 0.5\asec\\
\hline
\end{tabular}
\end{table*}

\subsection{Measurements and derived quantities}

\subsubsection{Rest-frame optical emission line fluxes and redshifts}

Measurements of rest-frame optical emission-line fluxes were obtained by fitting Gaussian profiles to the flux-calibrated one-dimensional MOSFIRE spectra.
The $1\sigma$ uncertainty on each line flux was estimated by perturbing the spectra $500$ times according to the error spectra, remeasuring the line flux, and taking the standard deviation of the resulting distribution.
The absolute flux calibration of the MOSFIRE spectra is accurate to within $18\%$ and the relative calibration between $H$-, and $K$-band filters is accurate to within 13$\%$ \citep{kriek2015}. These line flux measurements therefore provide robust estimates of absolute line luminosity (e.g., for estimating SFRs) and cross-filter line ratios (e.g., \oiiinwl/\oiinwl).
The same error simulations were used to determine line centroids and errors, and the associated redshift and redshift uncertainty for each emission line.
The individual line redshifts were combined using a weighted average to determine the final spectroscopic redshift and its associated uncertainty.

\subsubsection{Stellar masses and star-formation rates}\label{subsec:sedfit}

We estimated stellar masses using multiwavelength photometry from the VANDELS photometric catalogs \citep{mclure_vandels}, taking into consideration the fact that some photometric filters could be contaminated by rest-frame optical emission-line flux.
At high redshift ($z \gtrsim 2$), the large equivalent widths of the rest-frame optical emission lines \citep[e.g., typical rest-frame \oiiia \ equivalent-width values of $\geq 300${\AA} at $10^{10}$M$_{\odot}$;][]{reddy2018} can contaminate broadband photometry and result in an overestimation of galaxy masses.
In order to derive more accurate stellar masses, we first corrected the $H$-, and $K$-band photometry for the emission-line fluxes measured from the MOSFIRE spectra.
To perform this correction we constructed a model emission-line-only spectrum for each galaxy based on the line fits described above.
The flux contributed by emission lines to the $H$-, and $K$-band photometry was then determined by integrating the model over the appropriate filter profiles, and this flux was subtracted from the original photometry.
Corrections to the $H$- and $K$-band photometry ranged from $-0.3 < \Delta H / \rm{mag} < 0.0$ (median $\Delta H=-0.07$ mag) and $-1.8 < \Delta K / \rm{mag} < 0.0$ (median $\Delta K = -0.35$ mag) respectively.

The emission-line corrected photometry was modeled using \textsc{FAST++}\footnote{\href{https://github.com/cschreib/fastpp}{https://github.com/cschreib/fastpp}} \citep{schreiber2018}, an SED-fitting code closely based on the original FAST software \citep{kriek2009}.
We adopted \citet{conroy2009} flexible stellar population synthesis models and assumed solar metallicity\footnote{Although our analysis later in the paper (Section \ref{sec:results}) places the $\alpha$-element abundance of our sample at $\simeq0.3-0.5$ Z$_{\odot}$ and the Fe/H abundance at $\simeq0.1-0.2$ Z$_{\odot}$, adopting solar metallicity models is crucial for dust-correcting the nebular emission lines using the relation derived by \citet{sanders2020} (a different metallicity would bias the best-fitting continuum dust attenuation, see Secion \ref{subsec:nebular_dust_correction}).
This metallicity assumption will inevitably introduce some bias into the stellar mass determination \citep[see e.g.,][]{theios2019}, however it ensures our results and those of \citet{sanders2020} are directly comparable.}, a \citet{chabrier2003} initial mass function (IMF), constant star-formation histories and the \citet{calzetti2000} dust attenuation curve.
The redshift was fixed to the measured spectroscopic redshift.
These SED-fitting parameters were chosen to facilitate a direct comparison between our results and the recent study of gas-phase metallicities at $z\simeq3.3$ by \citet{sanders2020}.
The resulting fits yielded an estimate of the galaxy stellar mass (\mstar), star-formation rate (SFR$_{\mathrm{SED}}$), stellar continuum dust attenuation E(B-V)$_{\star}$) and a model of the stellar continuum.
Using the stellar continuum model we corrected the \hbeta \ line fluxes for underlying stellar absorption resulting in a median increase of $\simeq 3\%$ to the original flux values.

    \begin{figure}
        \centerline{\includegraphics[width=\columnwidth]{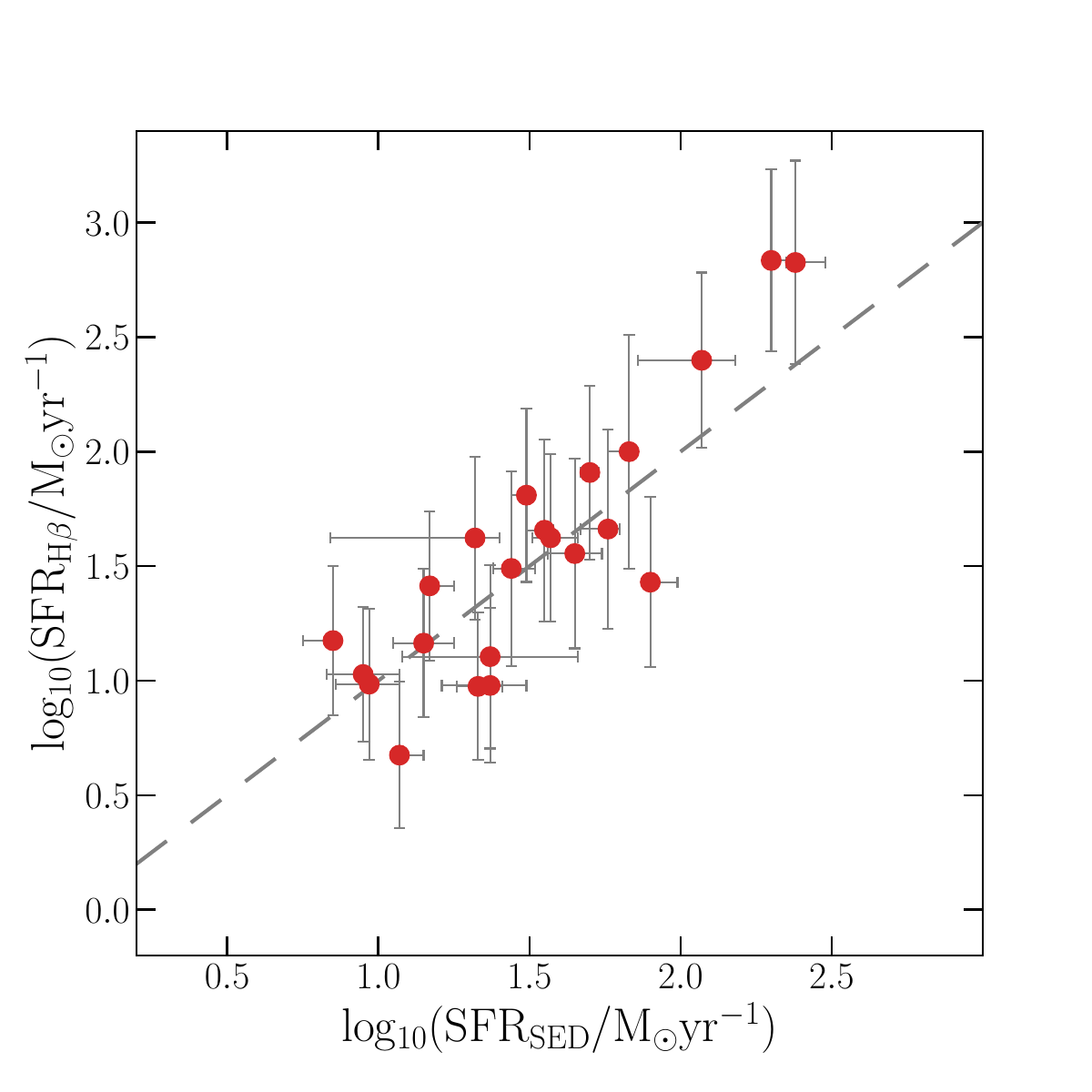}}
        \caption{A comparison between star-formation rate derived from the photometry and the dust-corrected \hbeta \ line flux for the 22 galaxies in our sample with a $>2\sigma$ detection of the \hbeta \ line. 
        The grey dashed line shows the one-to-one relationship.}
        \label{fig:sfrs}
    \end{figure}
    
\subsubsection{Dust-correcting nebular emission line fluxes}\label{subsec:nebular_dust_correction}
In order to determine accurate optical line ratios for estimating gas-phase metallicities, the observed line fluxes need to be corrected for nebular extinction.
Ideally, the nebular dust correction is determined directly using the Balmer decrement. 
However, this requires the detection of both the \halpha \ and \hbeta \ emission lines and is therefore not possible for our $z\simeq3.4$ sample.
Instead, we used the best-fitting value of the stellar continuum attenuation to dust-correct the emission-line fluxes for each galaxy.
We employed the calibration between stellar attenuation, SFR$_{\mathrm{SED}}$, redshift and nebular extinction described in \citet{sanders2020}, given by:

\begin{multline}\label{eq:newebvgas}
\text{E(B-V)}_{\text{neb}} = \text{E(B-V)}_{\text{stellar}} - 0.604 \\ + 0.538\times[\log(\text{SFR}_{\mathrm{SED}}) - 0.20\times(z - 2.3)] .
\end{multline}

\noindent
This calibration is based on observations of galaxies at $z\simeq2.3$ with both E(B-V)$_{\mathrm{neb}}$ (measured via the Balmer decrement) and E(B-V)$_{\star}$, and yields an unbiased estimate of E(B-V)$_{\rm{neb}}$ with an intrinsic scatter of 0.23 magnitudes and is directly applicable to our sample since we assume the same SED-fitting parameters used in the derivation of the calibration.
Based on the estimate of E(B-V)$_{\rm{neb}}$ given by equation \ref{eq:newebvgas}, the observed emission line fluxes were corrected for reddening assuming the \citet{cardelli1989} extinction law\footnote{As shown by \citet{reddy2020}, the nebular attenuation law in high-redshift star-forming galaxies closely follows the \citet{cardelli1989} Milky Way extinction curve.}.

To assess the reliability of the nebular dust correction we derived star-formation rates from the dust-corrected \hbeta \ line fluxes assuming an intrinsic ratio of \halpha/\hbeta$=2.86$ and applying the \citet{hao2011} \halpha-SFR conversion modified for a \citet{chabrier2003} IMF\footnote{To convert SFRs from the \citet{kroupa2001} IMF assumed in \citet{hao2011} to a \citet{chabrier2003} IMF we divide by 1.06.}.
In Fig. \ref{fig:sfrs} we show the resulting comparison between the \hbeta-derived SFRs and the original photometrically-derived SFR estimate for galaxies with a $>2\sigma$ detection of the \hbeta \ line.
The agreement is generally excellent, with a median offset of 0.06 dex in log(SFR) and a scatter of $\sigma=0.3$ dex, where $\sigma$ is derived from the median absolute deviation (MAD) ($\sigma=1.4826$ $\times$ $\mathrm{MAD}$).
    
    \begin{figure}
        \centering
        \centerline{\includegraphics[width=2.7in]{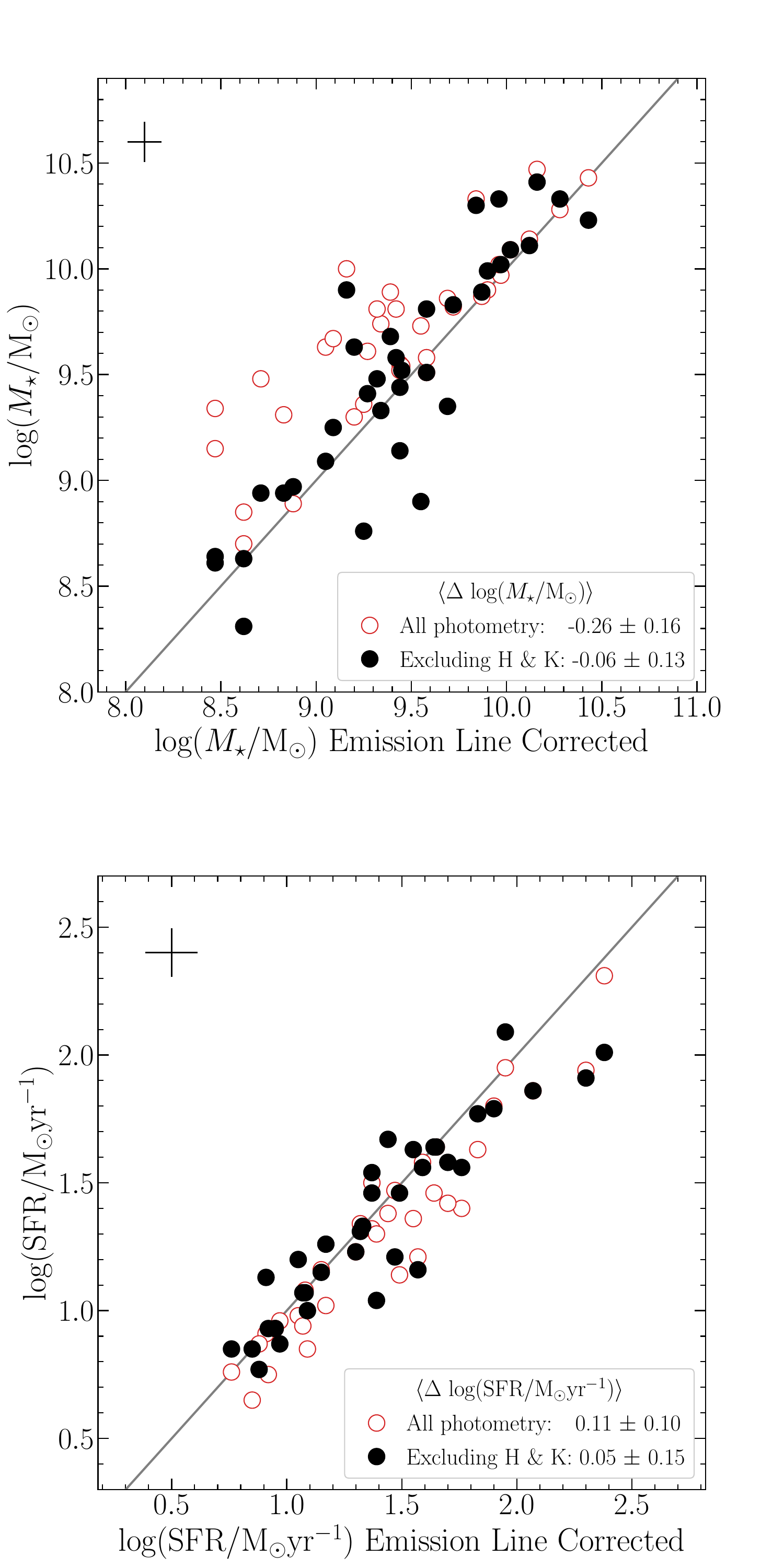}}
        \caption{The top panel shows SED-derived stellar masses after correcting the $H$- and $K$-band photometry for emission line contamination (see Section \ref{subsec:sedfit}) plotted against stellar masses derived from the original, non-corrected, photometry in which the $H$ and $K$ bands are included (open red circles) and excluded (filled black circles).
        The bottom panel shows the same for SED-derived star-formation rates.
        In each panel the grey solid line is the one-to-one relation and median error bars are shown in the top left-hand corner. 
        The mean, and standard deviation, of the offset from the one-to-one relation is indicated in the legend.
        The true, emission line corrected, values of \mstar and SFR can be reasonably recovered by simply excluding the contaminated $H$ and $K$ bands are excluded from the fit.}
        \label{fig:mass_sfr_bias}
    \end{figure}
    
\subsubsection{The effect of emission lines on SED-derived stellar masses and star-formation rates at $z\simeq3.4$}\label{subsec:sed_mass_offset}

In many instances, studies that rely on stellar mass estimates at $2.95 \leq z \leq 3.8$ do not have access to spectroscopic line flux measurements and as a result lack accurate corrections to the $H$- and $K$-band photometry.
It is therefore interesting to assess the effect of contamination from nebular emission lines on the derived stellar masses and SFRs at these redshifts.
In Fig. \ref{fig:mass_sfr_bias} we compare the values of \mstar \ and SFR(SED) estimated from corrected and uncorrected photometry.
If \mstar \ and SFR are derived using uncorrected $H$- and $K$-band photometry we find that, on average, log(\mstar) is overestimated by $0.26\pm0.16$ dex and log(SFR) is underestimated by $0.11\pm0.10$ dex.
These offsets are consistent with previous results in the literature.
In particular, the overestimation of \mstar \ is a well-known effect of the contamination of broad-band photometry by high-equivalent width rest-frame optical emission lines at high redshift \citep[e.g.,][]{schaerer2013, amorin2015, onodera2016}.
As can be seen from Fig. \ref{fig:mass_sfr_bias}, the effect is strongest at low \mstar \ where the equivalent widths are largest \citep[e.g.][]{reddy2018}.
Indeed, this bias due to optical line contamination is expected to be ubiquitous at $z>2$ due to the known evolution towards larger emission-line equivalent widths as a result of increasing specific star-formation rates and lower metallicity \citep{marmol-queralto2016,reddy2018}.

However, as can be seen from Fig. \ref{fig:mass_sfr_bias}, unbiased estimates of \mstar \ and SFR can be obtained by simply excluding the photometric bands that are known to include strong optical emission lines (in the case of $z\simeq3$ galaxies, the $H$- and $K$-bands).
With the contaminated bands excluded, the offsets in log(\mstar) and log(SFR) are consistent with zero ($-0.06\pm0.13$ dex and $0.05\pm0.15$ dex respectively).
In our particular case, the fact that we can recover these properties reliably in the absence of photometric anchors at rest-frame optical wavelengths (between $3000${\AA}$-6000${\AA}) is due to the data at longer wavelengths provided by the \emph{Spitzer} IRAC $3.6\mu$m and $4.5\mu$m imaging (covering rest-frame wavelengths between $8000${\AA}$-10000${\AA}).
In the absence of accurate line flux corrections, and with photometric data redward of the $K$-band, our results suggest that $-$ for the SED fitting assumptions described above $-$ simply excluding contaminated photometric bands from the SED-fitting process yields unbiased \mstar \ and SFR estimates that are consistent with the emission line-corrected values to within $\simeq10-15\%$.

\subsection{The $\mathbf{z\sim3.4}$ NIRVANDELS sample}

    \begin{figure*}
        \centerline{\includegraphics[width=7in]{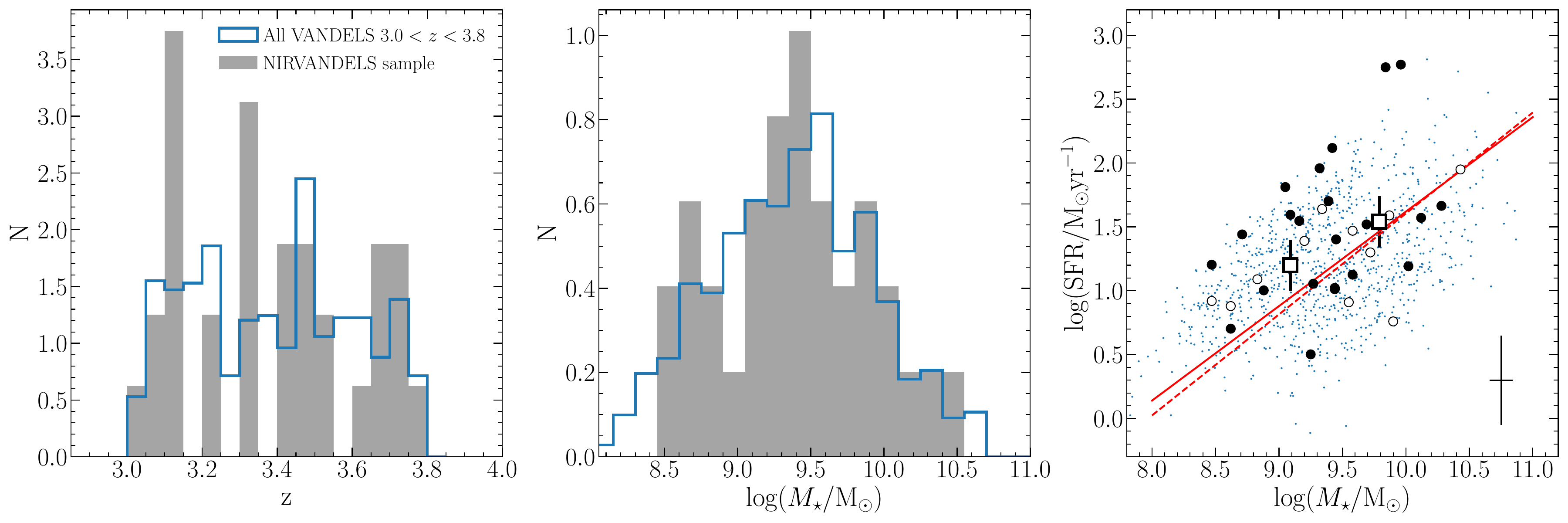}}
        \caption{The normalised redshift distribution (left), normalised stellar mass distribution (centre) and SFR - \mstar \ relation (right) for the star-forming galaxies in our NIRVANDELS sample.
        In the left-hand and centre panels, the open blue histogram represents all galaxies at $3.0<z<3.8$ in the VANDELS parent sample (N=980) and the filled grey histogram shows the subset of galaxies with MOSFIRE follow-up analysed here (N=33).
        In the right-hand panel the black filled circular points are galaxies with a \hbeta \ detection and therefore a SFR estimate based on the \hbeta \ flux; the open circular points represent galaxies without a \hbeta \ detection and therefore a SED-derived SFR estimate.
        The median uncertainty on \mstar \ and SFR is displayed in the lower right corner.
        The square data points show the high-\mstar \ and low-\mstar \ stacks (see Section \ref{subsec:stacks}).
        The small background data points shown the full VANDELS sample.
        The solid and dashed red lines show two paramterizations of the SFR-\mstar relation at $z=3.3$ derived by \citet{sanders2020} and \citet{speagle2014} respectively.
        }
        \label{fig:sample_properties}
    \end{figure*}

Our final sample was drawn from the 35 galaxies with spectroscopic redshifts in the range $3.0 \leq z \leq 3.8$ (Table \ref{tab:observations}).
We identified one galaxy in which the presence of active galactic nuclei (AGN) ionization was indicated by strong emission from the high-ionization species \nv, \civ \ and \heii \ in the rest-frame UV spectrum.
This galaxy was removed from our sample.
No further AGN were identified based on rest-frame UV and optical spectral properties.
Furthermore, for the remaining sample, we ruled out the presence of significant AGN ionization based on their mid-IR SED shapes and X-ray properties \citep[see][]{mclure_vandels}.
We also removed one galaxy in which the only optical emission feature detected was \oii.
There were two reasons for this decision.
Firstly, it is not possible to determine a gas-phase metallicity from the \oii \ doublet alone,  and therefore this object could not be included in our individual galaxy analysis.
Secondly, in our rest-frame optical stacking analysis (described in Section \ref{subsec:stacks} below) we required the detection of the \oiiia \ line in order to normalise the spectra, which was not available for this object.

Our final NIRVANDELS sample therefore consisted of 33 galaxies with secure spectroscopic redshifts in the range $2.95 \leq z \leq 3.8$.
Each object in the sample has MOSFIRE spectra in the $H$- and $K$-bands covering a number of rest-frame optical emission lines sufficient for deriving gas-phase metallicity (\zgas), and a VIMOS/VANDELS rest-frame FUV spectrum from which the stellar metallicity (\zstar) can be determined.

In Fig. \ref{fig:sample_properties} we show the normalised redshift and mass distributions of the NIRVANDELS galaxies compared to the full VANDELS sample of $\mathrm{N}=791$ star-forming galaxies in the redshift range $3.0 \leq z \leq 3.8$.
The stellar masses of the full sample have been derived using the same SED fitting procedure described in Section \ref{subsec:sedfit}.
As the photometry for the full sample cannot be corrected for emission-line contamination, we have excluded the $H$- and $K$-band photometry from these fits for the reasons described in Section \ref{subsec:sed_mass_offset} (all objects benefit from longer-wavelength \emph{Spitzer} IRAC data).
The properties of the sample with respect to the full $z\simeq3.4$ star-forming galaxy population are discussed further in Section \ref{subsec:representative_sample}.

\subsubsection{Composite spectra}\label{subsec:stacks}

Our study is focused on determining stellar metallicities (\zstar) and gas-phase metallicities (\zgas), and, wherever possible, we estimated these quantities on an individual galaxy basis.
However, in order to include galaxies for which such measurements were not possible, we also made use of stacked spectra.
In order for an object to be included in the stacking sample, we required coverage of all rest-frame optical emission lines in the MOSFIRE spectra (\oii, \neiii, \hbeta \ and \oiii). 
In total, 5 objects were removed from the stacking sample due to fact that the \hbeta \ line was not covered by the detector
\footnote{We note that these five galaxies were detected in \oiiia \ and \oii, allowing \zgas \ to be determined on an individual basis using the \oiiia/\oii \ ratio (see Section \ref{sec:metallicity_derivation}). However, including these objects in the stacks would clearly bias the \hbeta-dependent line ratios and \hbeta-derived average SFRs.}.
The final stacking sample therefore contained only 28/33 galaxies, and is fully representative of the complete sample.

In this paper, we focus on the correlation between metallicity and stellar mass, and we therefore constructed stacks in two bins of \mstar: high-\mstar \ and low-\mstar, split at the median \mstar \ of the stacking sample (\mstar \ $=10^{9.4}$M$_{\odot}$).
We employed the method described in \citet{cullen2019} to produce FUV stacks from the VANDELS data and refer readers to that paper for further details.
For the MOSFIRE spectra, we first converted each spectrum into luminosity density units using its spectroscopic redshift and corrected for nebular dust attenuation using E(B-V)$_{\rm{neb}}$ assuming the \citet{cardelli1989} Milky Way extinction curve.
To avoid biasing the stacks in favour of the brightest objects we normalised each spectrum using its measured \oiiia \ luminosity\footnote{We note that this choice has the potential to introduce subtle biases due to the exact details of the sample, and the metallicity dependence of the chosen line etc. However, we have confirmed that this choice does not have a strong affect on the measured line ratios and derived gas-phase metallicities; combining the galaxies without this normalisation does not change the results of this paper.}.
The normalised spectra were then re-sampled onto a common rest-frame wavelength grid, and the final stacked spectrum was produced by taking the median value at each wavelength.
The error at each wavelength was calculated via a bootstrap re-sampling of the individual values. 
Finally, the normalised stacks were converted back into absolute luminosity density units by multiplying by the median \oiiia \ luminosity of the contributing galaxies.

\renewcommand{\arraystretch}{1.5}
\begin{table}
    \centering
    \caption{Properties of the individual galaxies in the NIRVANDELS sample, including the derived stellar and gas-phase metallicities.}\label{table:observed_properties}
    \begin{tabular}{ccrcc}
        \hline
        Name & $z_{\mathrm{neb}}$ &  log($M_{\star}$/M$_{\odot}$) & log($Z_{\mathrm{g}}/\mathrm{Z}_{\odot}$) & log($Z_{\star}/\mathrm{Z}_{\odot}$) \\
        \hline
        \hline
        NIRV-25444 & 3.7019 & 9.87      & $-0.29^{+0.26}_{-0.22}$ & $\cdots$ \\
        NIRV-25568 & 3.2109 & 10.12     & $\cdots$ & $\cdots$ \\
        NIRV-25732 & 3.1447 & 8.83      & $-0.49^{+0.12}_{-0.13}$ & $\cdots$ \\
        NIRV-28864 & 3.5172 & 9.55      & $\cdots$ & $\cdots$ \\
        NIRV-29419 & 3.7048 & 10.28     & $-0.24^{+0.15}_{-0.19}$ & $\cdots$ \\
        NIRV-30119 & 3.7713 & 8.88      & $-0.68^{+0.09}_{-0.09}$ & $\cdots$ \\
        NIRV-30602 & 2.9832 & 9.72      & $-0.28^{+0.22}_{-0.21}$ & $\cdots$ \\
        NIRV-30845 & 3.1452 & 9.69      & $-0.43^{+0.21}_{-0.24}$ & $\cdots$ \\
        NIRV-31538 & 3.3131 & 10.02     & $\cdots$ & $\cdots$ \\
        NIRV-31982 & 3.4702 & 9.44      & $-0.45^{+0.13}_{-0.10}$ & $\cdots$ \\
        NIRV-33568 & 3.3085 & 8.47      & $-0.77^{+0.06}_{-0.08}$ & $\cdots$ \\
        NIRV-33613 & 3.7076 & 9.58      & $-0.36^{+0.24}_{-0.22}$ & $\cdots$ \\
        NIRV-33644 & 3.2049 & 9.32      & $-0.44^{+0.15}_{-0.12}$ & $-0.93^{+0.22}_{-0.12}$ \\
        NIRV-34030 & 3.0750 & 9.90      & $\cdots$ & $\cdots$ \\
        NIRV-34438 & 3.1383 & 9.09      & $\cdots$ & $-0.93^{+0.36}_{-0.20}$ \\
        NIRV-34449 & 3.4714 & 9.58      & $-0.45^{+0.23}_{-0.26}$ & $\cdots$ \\
        NIRV-34500 & 3.6008 & 9.42      & $-0.44^{+0.17}_{-0.19}$ & $\cdots$ \\
        NIRV-34591 & 3.3229 & 9.25      & $\cdots$ & $\cdots$ \\
        NIRV-34777 & 3.4042 & 9.39      & $-0.59^{+0.18}_{-0.16}$ & $-0.73^{+0.26}_{-0.31}$ \\
        NIRV-34889 & 3.6712 & 9.96      & $-0.01^{+0.16}_{-0.19}$ & $\cdots$ \\
        NIRV-35212 & 3.4024 & 9.84      & $-0.38^{+0.17}_{-0.19}$ & $-0.54^{+0.10}_{-0.10}$ \\
        NIRV-35557 & 3.3470 & 9.27      & $-0.47^{+0.09}_{-0.07}$ & $\cdots$ \\
        NIRV-35865 & 3.4017 & 10.43     & $\cdots$ & $\cdots$ \\
        NIRV-35915 & 3.1159 & 8.62      & $\cdots$ & $\cdots$ \\
        NIRV-36951 & 3.0668 & 8.71      & $-0.56^{+0.06}_{-0.06}$ & $\cdots$ \\
        NIRV-37402 & 3.1360 & 8.47      & $\cdots$ & $\cdots$ \\
        NIRV-38315 & 3.4986 & 9.05      & $-0.72^{+0.09}_{-0.09}$ & $\cdots$ \\
        NIRV-38451 & 3.5091 & 9.16      & $-0.55^{+0.08}_{-0.09}$ & $\cdots$ \\
        NIRV-41105 & 3.0039 & 8.62      & $\cdots$ & $\cdots$ \\
        NIRV-42161 & 3.1113 & 9.34      & $\cdots$ & $-0.77^{+0.21}_{-0.18}$ \\
        NIRV-44246 & 3.6770 & 9.44      & $-0.62^{+0.09}_{-0.08}$ & $\cdots$ \\
        NIRV-45896 & 3.6747 & 9.20      & $\cdots$ & $\cdots$ \\
        NIRV-46857 & 3.3490 & 9.45      & $-0.44^{+0.16}_{-0.16}$ & $-0.72^{+0.18}_{-0.18}$ \\
        \hline
    \end{tabular}
\end{table}

Due to the fact that the optical emission line profiles in the stacked spectra were not necessarily well-described by a single Gaussian profile, line luminosities for the stacks were determined via direct integration after subtracting off the local stellar continuum.
The stacked \hbeta \ line luminosities were corrected for stellar absorption using the median correction of the galaxies contributing to the stack.
These correction factors were relatively small, at the $\simeq 3\%$ level.

To estimate uncertainties on the various properties derived from the stacked spectra (line ratios/luminosities, \zstar, \zgas) we employed a re-sampling methodology.
The \mstar \ values of galaxies in the stacking sample were perturbed according to their $1\sigma$ uncertainties and the high-\mstar \ and low-\mstar \ bins re-populated with replacement.
The E(B-V)$_{\rm{neb}}$ values for each galaxy were also perturbed according to their $1\sigma$ uncertainties.
Finally, the 1D spectra were perturbed using their error spectra.
The stacks were then re-constructed using the methodology described above and the various quantities of interest were remeasured. 
This process was repeated $500$ times.  
Uncertainties on all quantities measured from stacked spectra were then calculated using the $16$th and $84$th percentiles of the resulting distributions.

    \begin{figure*}
        \centerline{\includegraphics[width=6.6in]{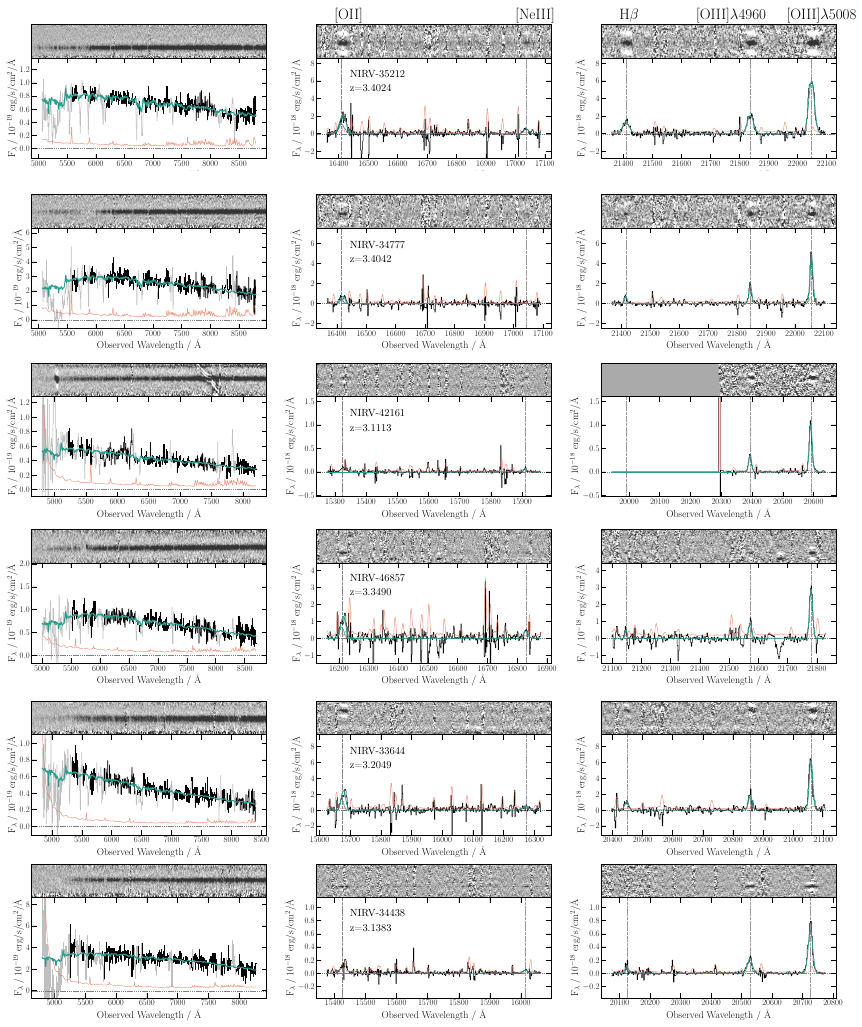}}
        \caption{From left to right: VANDELS rest-frame FUV spectrum, MOSFIRE $H$-band spectrum, and MOSFIRE $K$-band spectrum for the $\mathrm{N}=6/33$ individual galaxies in our sample for which were able to estimate \zstar. These galaxies have a median SNR per resolution element $>= 5$ in the VANDELS rest-frame FUV spectrum.
        In each panel, both 1D and 2D spectra are shown in the lower and upper portion of the panel, respectively.
        For the 1D spectra, the observed spectrum is shown in black, with the best-fitting model over-plotted in blue and the error spectrum shown in orange.
        For the VANDELS rest-frame FUV 1D spectra, regions of the spectrum dominated by ISM absorption lines/nebular emission lines are masked out (lighter shading) as these wavelength regions are not included in the stellar continuum fits used to determine \zstar.
        In the MOSFIRE panels, dotted vertical lines indicate the positions of the nebular emission lines used to determine \zgas \ (\oiinwl, \neiiinwl, \hbeta, \oiiinwl); the line labels are given at the top of the figure.
        }
        \label{fig:individual_fits}
    \end{figure*}

\subsubsection{A representative sample?}\label{subsec:representative_sample}

If the results of this study are to be applied in general to star-forming galaxies at $z\simeq3.4$, it is necessary to demonstrate that our sample is not a highly-biased subset of the general galaxy population at this epoch.
In particular, when considering the mass-metallicity relation, it is important to assess whether the galaxy sample is biased in terms of SFR at fixed \mstar.
Numerous studies in the literature, both at $z=0$ and higher redshifts, have found strong evidence for an anti-correlation between SFR and metallicity at fixed \mstar \ \citep[i.e, the FMR;][]{mannucci2010,sanders2018}.
Therefore, in order to avoid SFR biases, the galaxies in our sample should be representative of the `star-forming main sequence' (SFR-\mstar relation) at $z\simeq3.4$.

In the right-hand panel of Fig. \ref{fig:sample_properties} we show the location of the NIRVANDELS galaxies on the SFR-\mstar plane compared to two literature determinations of the SFR-\mstar relation at $z\simeq3.4$ \citep{speagle2014, sanders2020}.
We also show the underlying distribution of all VANDELS galaxies at $2.95 \leq z \leq 3.80$ to illustrate the typical range of SFRs observed at fixed \mstar.
Above \mstar$\simeq10^{9.4}$\msun (the median stellar mass of our sample), the scatter of the individual NIRVANDELS galaxies about the main sequence is consistent with that of the underlying population and there is no evidence that our sample is strongly biased in terms of SFR at these stellar masses.
Moreover, the high-\mstar \ stack is consistent with the \citet{speagle2014} main sequence relation within $1\sigma$. 
Therefore, for galaxies with \mstar \ $\gtrsim10^{9.4}$ \msun \ (our high-\mstar \ sample), we conclude that the results presented in this paper are applicable to the general $z\simeq3.4$ star-forming population.

At \mstar \ $<10^{9.4}$ \msun, the individual NIRVANDELS galaxies predominantly fall above the literature relations. 
The low-\mstar \ stack also sits above the main sequence by $\simeq0.3$ dex.
Although the stack is formally consistent with the \citet{speagle2014} relation within $2\sigma$, we clearly cannot rule out the possibility that our sample is biased towards high SFR at the lowest stellar masses.
This in turn could lead to a bias in the MZR.
Based on the results of \citet{sanders2018} the \zgas \ bias at fixed \mstar \ due to an offset from the main-sequence can be estimated as $\Delta$\zgas$\simeq -0.15 \times \Delta \mathrm{log(SFR)}$.
An offset of $0.3$ dex therefore represents a bias of $\simeq -0.05$ dex in \zgas\footnote{We note that the offset is slightly larger ($\simeq -0.08$ dex) assuming another recent determination of the FMR presented in \citet{curti2020}.}.
We chose not to correct for this potential bias in this paper because (i) the magnitude of the offset is similar to the typical uncertainties on our individual \zgas \ measurements; and  (ii) the main sequence offset in itself is not highly significant ($<2\sigma$).
Nevertheless, we caution that at \mstar \ $\lesssim10^{9.4}$ \msun \ our sample may include a small bias towards objects with elevated SFRs and lower metallicities.

\renewcommand{\arraystretch}{1.5}
\begin{table*}
    \centering
    \caption{Properties of the low-\mstar \ and high-\mstar \ composites.}\label{table:stacked_properties}
    \begin{tabular}{rrcccccc}
        \hline
        Stack & log($M_{\star}$/M$_{\odot}$) Range & Median log($M_{\star}$/M$_{\odot}$) & \oiiinwl/\hbeta & \oiiinwl/\oiinwl & \neiiinwl/\oiinwl & log($Z_{\mathrm{g}}/\mathrm{Z}_{\odot}$) & log($Z_{\star}/\mathrm{Z}_{\odot}$) \\
        \hline
        \hline
        low-\mstar & $8.50-9.40$ & 9.09 & $0.88\pm0.06$ & $0.55\pm0.10$ & $-0.64\pm0.15$ & $-0.54 \pm 0.06$ & $-0.89\pm0.06$ \\
        high-\mstar & $9.40-10.40$ & 9.81 & $0.58\pm0.06$ & $0.22\pm0.09$ & $-0.76\pm0.19$ & $-0.31\pm0.06$ & $-0.72\pm0.07$\\
        \hline
    \end{tabular}
\end{table*}

\section{Determination of galaxy metallicities}\label{sec:metallicity_derivation}

The primary focus of this study is the determination of gas-phase metallicities (\zgas) and stellar metallicities (\zstar)  as a function of galaxy stellar mass at $z\simeq3.4$.
Below we describe in detail how each of these parameters was measured from the spectroscopic data. 

Before going into the details of these methods, it is worth again emphasizing that \zgas \ and \zstar, as measured in this paper, are not sensitive to the same element abundances (see also Section \ref{sec:intro}).
Gas-phase metallicities, determined from rest-frame optical nebular emission lines, are sensitive to the elements that act to cool the $T\simeq10^4$K ionized gas, namely oxygen (O/H).
On the other hand, stellar metallicities derived from rest-frame FUV spectra are sensitive to the elements that dominate the FUV opacity of massive stars, namely iron (Fe/H) \citep{leitherer2010}.
We will discuss the implications of this difference in detail in Section \ref{sec:ofe_results}.

\subsection{Determination of the gas metallicity ($\mathbf{Z_{\rm{g}}}$)}\label{subsec:zgas_meaurement}

Estimates of the gas-phase metallicity (\zgas, or O/H) were derived using the ratios of rest-frame optical emission lines measured in the MOSFIRE spectra.
In the redshift range investigated in this paper ($2.95 \leq z \leq 3.80$) the \oii \ and \neiii \ lines are available in the $H$-band, while the \hbeta \ and \oiii \ lines are available in the $K$-band.
We selected three independent metallicity diagnostics from these lines: \oiiia/\hbeta, \oiii/\oii \ (O$_{32}$) and \neiii/\oii.
To estimate \zgas \ from these ratios we followed the method described in \citet{sanders2020} using the high-redshift analogue calibrations of \citet{bian2018}.
These calibrations are known to be consistent with the (currently small) sample of $z\simeq2$ galaxies for which direct-method O/H estimates are available \citep{sanders2020a}.
Using these calibrations, we calculated the best-fitting value of \zgas \ via a $\chi^2$ minimization using:
\begin{equation}\label{eq_zgas_est}
\chi^2(x)=\sum_{i}^{}\frac{(\mathrm{R}_{\mathrm{obs,i}}-\mathrm{R}_{\mathrm{cal,i}}(x))^2}{(\sigma_{\mathrm{obs,i}}^2+\sigma_{\mathrm{cal,i}}^2)}.
\end{equation}
where $x=\mathrm{12+log(O/H)}=\mathrm{log}(Z_{\mathrm{g}}/\mathrm{Z}_{\odot})+8.69$, the sum over $i$ represents the line ratios used, $\mathrm{R}_{\mathrm{obs,i}}$ is the logarithm of the  $i$-th observed line ratio, $\mathrm{R}_{\mathrm{cal,i}}(x)$ is the predicted value of R$_{i}$ at $x$ from the \citet{bian2018} calibrations, $\sigma_{\mathrm{obs,i}}$ is the uncertainty on $\mathrm{R}_{\mathrm{obs,i}}$, and $\sigma_{\mathrm{cal,i}}$ is the calibration uncertainty.
The best-fitting \zgas \ solution is found by minimizing the value of $\chi^2$ in equation \ref{eq_zgas_est}.
Uncertainties were estimated by perturbing the observed line ratios by their $1\sigma$ error values and re-calculating \zgas \ 500 times.
The $1\sigma$ uncertainty on \zgas \ was derived from the 68th percentile width of the resulting distribution.

In practice, not all lines were detected for a given object.
In this case, the minimum requirement for a metallicity estimate was the detection of the \oiiia \ and \oii \ lines. It is possible to estimate \zgas \ using only the O$_{32}$ ratio because the O$_{32}$-\zgas \ calibration is monotonic.
The \oiiia/\hbeta \ calibration, on the other hand, is double valued and therefore cannot be used without another ratio to break the degeneracy. 
The \neiii/\oii \ ratio is also monotonic and could, in principle, be used to infer metallicity \citep[e.g.,][]{shapley2017} but in practice there were no instances in which this was the only line ratio available\footnote{This is due to the fact that \neiii \ is the faintest line considered in this paper.
All galaxies with a \neiii \ detection were also detected in \oiiia \ and \oii.}.

    \begin{figure}
        \centerline{\includegraphics[width=\columnwidth]{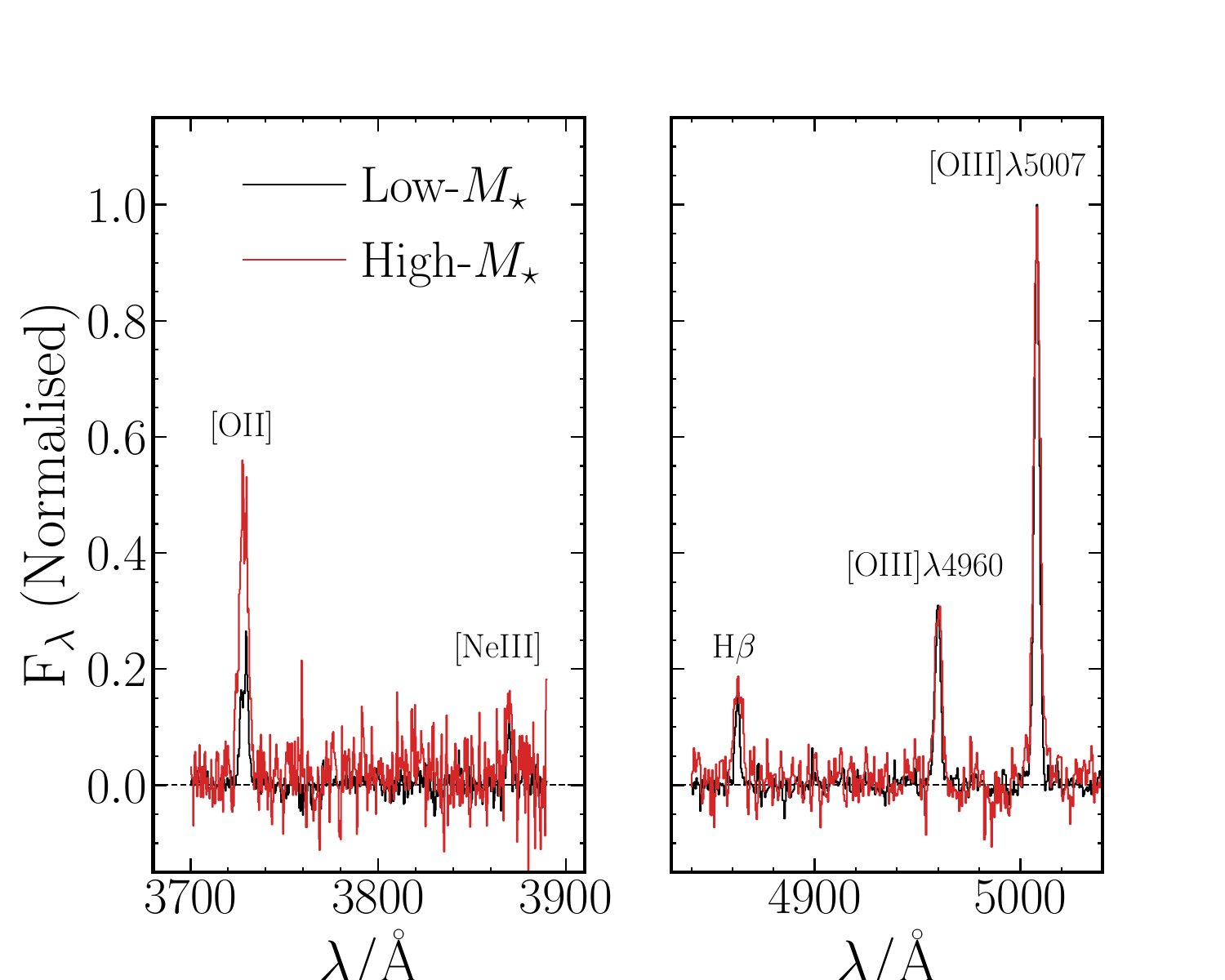}}
        \caption{Portions of the MOSFIRE composite spectra in the $H$ band (left) and $K$ band (right) covering the \oiinwl, \neiiinwl, \hbeta \ and \oiiinwl \ nebular emission lines.
        The low-\mstar \ composite is shown in black and the high-\mstar \ stack in red.
        Both composites have been normalized by the peak \oiiia \ flux of the low-\mstar \ composite spectrum.
        }
        \label{fig:opt_stacks}
    \end{figure}


    \begin{figure*}
        \centerline{\includegraphics[width=7.8in]{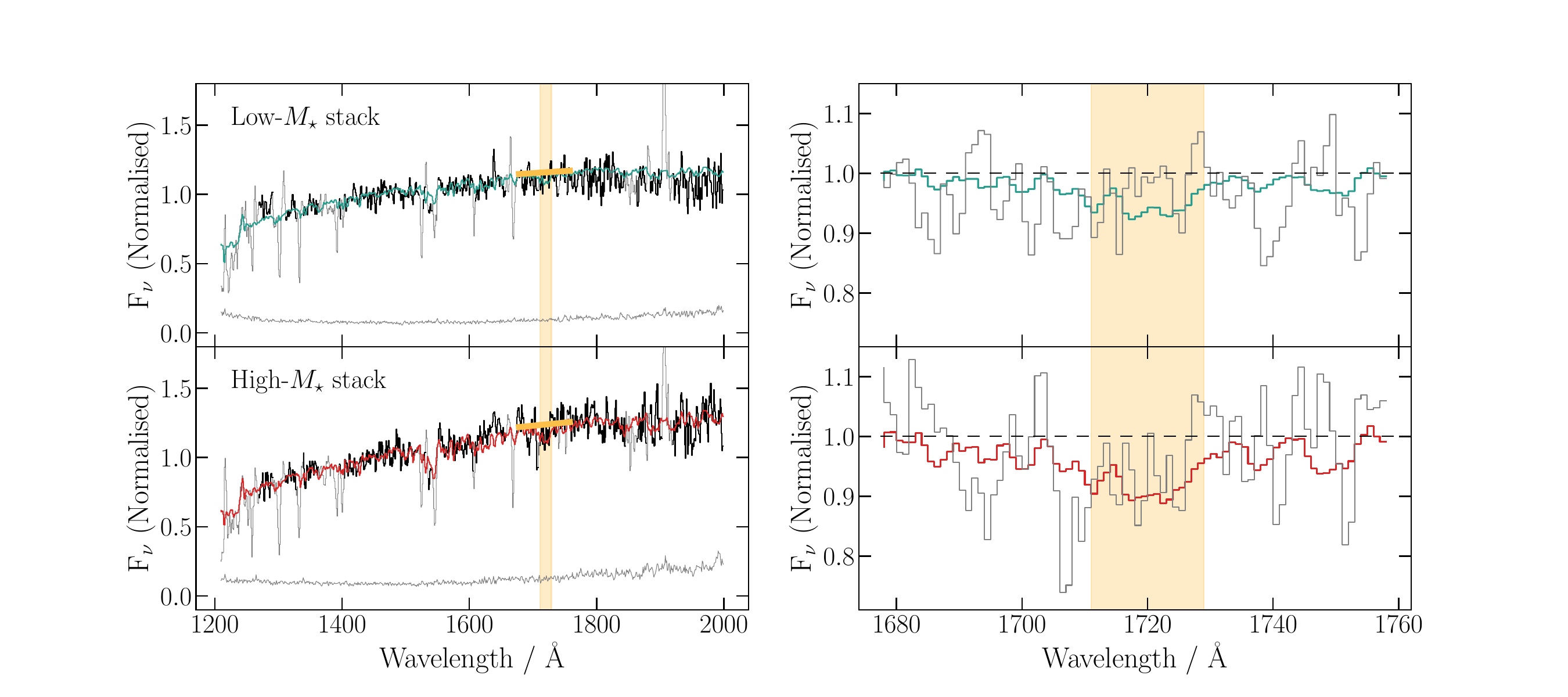}}
        \caption{Stacked FUV composite spectra for the low-\mstar \ (top panels) and high-\mstar \ (bottom panels) sample.
        The left-hand panels show the full FUV spectrum in the range $1200-2000$ {\AA}, normalised by the median flux in the range $1420-1480$ {\AA}.
        Bootstrapped error spectra are shown in grey.
        The best-fitting Starburst99 models are superposed on top of the composites (in turquoise and red for the low-\mstar \ the high-\mstar \ composites, respectively).
        Regions of the composite spectra in a lighter shade highlight the position of strong interstellar absorption features or nebular emission lines that are not included in the model fitting.
        The yellow shaded region shows the position of the \zstar-sensitive `1719' index \citep[e.g.][]{calabro2021}.
        The right-hand panels show continuum-normalised versions of both spectra in the region of the `1719' index.
        The continuum normalisation was performed using the surrounding \citet{rix2004} windows and the level of the continuum is shown is shown by the thick yellow line in both of the left-hand panels.
        It can be seen that the high-\mstar \ composite shows stronger absorption across the `1719' index, indicating higher \zstar \ \citep[][]{calabro2021}.
        }
        \label{fig:fuv_stacks}
    \end{figure*}
    
For each galaxy we utilised the maximum number of line ratios possible, which resulted in $6/33$ metallicities determined using all three line ratios, $11/33$ using  O$_{32}$ and \oiiia/\hbeta, and $4/33$ using O$_{32}$ alone.
Using more line ratios improves the constraints on \zgas \ but does not bias the solution on average \citep{sanders2020}.
The remaining $12/33$ galaxies do not have individual \zgas \ determinations; in all cases this was due to a non-detection of the \oii \ line.
For both the low-\mstar \ and high-\mstar \ composite spectra all lines were detected.
The gas-phase metallicities estimates resulting from this procedure are listed in Tables \ref{table:observed_properties} and \ref{table:stacked_properties}.
Examples of the nebular emission line spectra for individual objects are shown Fig. \ref{fig:individual_fits} and a comparison between the two composite spectra are shown in Fig. \ref{fig:opt_stacks}.

    \begin{figure*}
        \centerline{\includegraphics[width=7in]{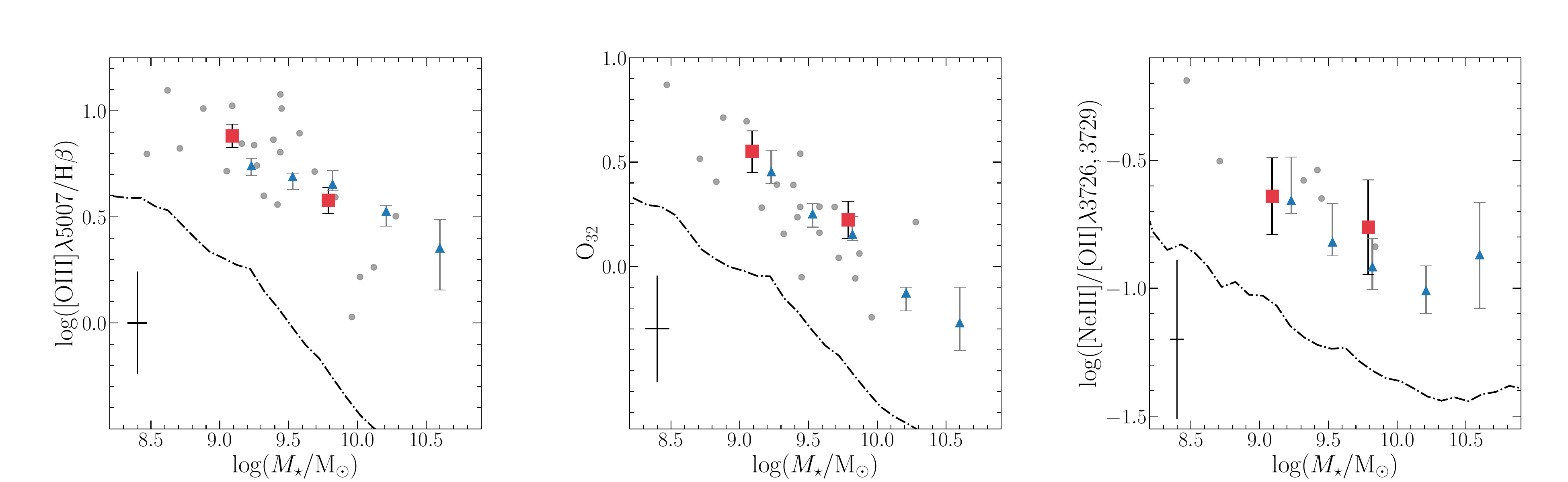}}
        \caption{Dust-corrected emission lines ratios versus \mstar \ for galaxies at $z\simeq3.4$ and $z\simeq0$.
        Each panel shows one of the three emission line ratios used to determine \zgas \ in this paper (see Section \ref{sec:metallicity_derivation}).
        From left to right these line ratios are: \oiiinwl/\hbeta, O$_{32}$, and \neiiinwl/\oiinwl.
        The red squares show the low-\mstar \ and high-\mstar \ NIRVANDELS stacks (see also Fig. \ref{fig:opt_stacks}) and the grey circles show the individual NIRVANDELS galaxies with the necessary line detections.
        The median error bar on the individual detections is shown in the bottom left-hand corner.
        The blue triangular data points show stacked measurements from the $z\simeq3.3$ MOSDEF sample presented in \citet{sanders2020}.
        The dot-dashed lines represent the running medians for $z\simeq0$ galaxies from the SDSS sample of \citet{andrews2013}.
        These $z\simeq0$ line ratios have been corrected, where possible, for diffuse ionized gas (DIG) contamination that can bias line flux measurements in local galaxies (see text and \citet{sanders2020} for further details).}
        \label{fig:emission_line_properties}
    \end{figure*}

\subsection{Determination of the stellar metallicity ($\mathbf{Z_{\star}}$)}\label{subsec:zstar_meaurement}

Stellar metallicities (\zstar, or Fe/H) were estimated from the rest-frame FUV spectra using a full spectral fitting technique within the wavelength range $1221-2000$ {\AA} as described in \citet{cullen2019, cullen2020}.
The method involves fitting stellar population synthesis models to all portions of the FUV spectra dominated by stellar continuum emission (avoiding regions contaminated by ISM absorption lines and/or nebular emission lines).
The metallicity of the model (\zstar) is constrained while marginalising over three nuisance parameters related to dust attenuation.
Below we briefly describe this technique but refer readers to \citet{cullen2019} for full details.

For the stellar population models we adopted the Starburst99 (SB99) high-resolution WM-Basic models described in \citet{leitherer2010}.
To construct the models we assumed constant star formation over timescales of 100 \Myr \ and adopted the weaker-wind Geneva tracks with stellar rotation and single-star evolution at the following metallicities: $Z_{\star}=(0.001, 0.002, 0.008, 0.014, 0.040)$.
The models were fitted using the nested sampling algorithm implemented in the  python package \texttt{dynesty} \citep{speagle2020}.
The free parameters in the fit were \zstar \ and three nuisance parameters that define the shape of the FUV attenuation curve using the parameterisation described in \citet{salim2018}.
As the models are provided at five fixed \zstar \ values, we linearly interpolated the logarithmic flux values between the models in order to generate a model for any \zstar \ value within the prescribed range.
The models were then convolved to the resolution of the VANDELS spectra and appropriately re-sampled.

We used a log-likelihood function of the form,
\begin{align}
\begin{split}
\mathrm{ln}(L) &= K-\frac{1}{2}\sum_{i}\bigg[\frac{(f_{i}-f(\theta)_{i})^2}{\sigma_{i}^2}\bigg] \\
&=K-\frac{1}{2}\chi^2
\end{split}
\end{align}
where $K$ is a constant, $f$ is the observed flux, $f(\theta)$ is the model flux for a given set of parameters $\theta$, and $\sigma$ is the error on the observed flux.
The likelihood was computed using only those wavelength pixels free from ISM absorption or nebular emission-line contamination.
For this purpose we adopted the `Mask 1' windows defined in \citet{steidel2016}.
The \texttt{dynesty} algorithm provides estimates of the posterior probability distributions for each of the free parameters in the fit.
For a given fit, the best-fitting \zstar \ value was calculated from the $50$th percentile of the resulting \zstar \ posterior distribution and the uncertainty derived from the 68th percentile width.


Deriving stellar metallicities for individual galaxies is generally more difficult than deriving gas-phase metallicities due to the requirement for a high S/N detection of the continuum.
In this paper, we only report \zstar \ values for the six individual galaxies for which the average S/N per resolution element of the VANDELS spectrum is $\geq 5$ in the relevant wavelength range.
At lower S/N, we find that the typical $1\sigma$ uncertainties on \zstar \ are $>50\%$.
Moreover, below a certain threshold S/N estimates of \zstar \ can become biased. 
For example, \citet{topping2020b} find that unbiased estimates of \zstar \ require S/N $\geq 5.6$ per resolution element (five of the six galaxies with individual \zstar \ estimates satisfy this slightly higher threshold).
Stellar metallicity estimates resulting from this procedure are listed in Tables \ref{table:observed_properties} and \ref{table:stacked_properties}, and examples of fits to the individual and stacked FUV  spectra are shown in Figs. \ref{fig:individual_fits} and \ref{fig:fuv_stacks}.
In Fig. \ref{fig:fuv_stacks} we also show a zoom-in of continuum-normalised versions of the stacks in the region of the `1719' index, recently advocated by \citet{calabro2021} as a useful \zstar-sensitive index in the FUV.
In agreement with the results of \citet{calabro2021}, we find that the higher metallicity, high-\mstar, stack shows increased absorption in the `1719' index region.
    
\section{Results}\label{sec:results}


\subsection{The gas-phase mass-metallicity relation at $\mathbf{z\simeq3.4}$}

\subsubsection{Trends with emission-line ratios}\label{subsec:emline_ratios}

Since the gas-phase metallicity is estimated directly from emission-line ratios, we first examine the empirical trends between the observed dust-corrected line ratios, stellar mass and redshift.
In Fig. \ref{fig:emission_line_properties}, we show the three emission-line ratios used to estimate \zgas \ (\oiiinwl/\hbeta, O$_{32}$ and \neiiinwl/\oiinwl) as a function of \mstar \ for our sample.
Also shown is an independent dataset of star-forming galaxies at $z\simeq3.3$ from \citet{sanders2020}, and a sample of local star-forming galaxies drawn from the Sloan Digital Sky Survey (SDSS).
The SDSS galaxies are taken from the \citet{andrews2013} sample as described in \citet{sanders2020}.
To facilitate a direct comparison with our $z\simeq3.4$ galaxies, the SDSS line ratios have been corrected, where possible, for diffuse ionized gas (DIG) emission\footnote{For the \oiiinwl/\hbeta \ and \oiiinwl/\oiinwl \ ratios, the SDSS data are DIG-corrected, however this is not possible for \neiiinwl/\oiinwl \ due to the lack of an accurate DIG-correction for the \neiiinwl \ line \citep{sanders2020}.}.
DIG is not associated with \hii \ regions and therefore biases nebular emission line flux measurements in the integrated spectra of galaxies at $z\simeq0$ \citep{oey2007,sanders2017,valeasari2019}.
A DIG correction is not required for the $z\simeq3.4$ sample because the DIG contribution becomes negligible at the typical star-formation surface densities of high-redshift galaxies \citep{sanders2017,shapley2019}.
    
Fig. \ref{fig:emission_line_properties} clearly demonstrates that each of the three line ratios increases with decreasing stellar mass.
The negative correlation between line ratio and \mstar \ applies to both the galaxies at $z\simeq3.4$ and the local SDSS sample.
Based on the assumption that the galaxies lie on the upper-branch of the \oiiinwl/\hbeta-\zgas \ relation \footnote{We can assert this with confidence based on the observed \oiiinwl/\oiinwl \ ratios, which rule out galaxies lying on the lower branch of the \oiiinwl/\hbeta-\zgas \ relation assuming the \citet{bian2018} calibration is applicable to our sample \citep[see also][]{curti2020}.}, these trends indicate that \zgas \ increases with increasing \mstar.
Crucially, this correlation holds irrespective of exactly which emission-line calibration is adopted \citep[e.g.][]{maiolino2008,bian2018,kewley2019,curti2020}.
For our $z\simeq3.4$ sample, the trend is clearly evident in the \oiiinwl/\hbeta-\mstar \ and O$_{32}$-\mstar \ diagrams for both the individual galaxies and stacks, and is fully consistent with the \citet{sanders2020} data.
The consistency with the \citet{sanders2020} line ratios, which are based on a much larger sample of $\mathrm{N}=245$ galaxies drawn from the star-forming main sequence, provides further evidence that our sample is not a highly-biased subset of the star-forming galaxy population at $z\simeq3.4$.
As an aside, we also note that the position of our sample in the \oiiinwl/\hbeta-\mstar \ plane overlaps with the star-forming galaxy region of the Mass-Excitation (MEx) diagnostic diagram proposed by \citet{juneau2014}, providing further evidence that our sample is not strongly contaminated by AGN excitation.

For the \neiiinwl/\oiinwl \ ratio the situation is less clear due to the difficulty in detecting the faint \neiiinwl \ line.
Nevertheless, for the small number of individual detections, a trend is apparent that is consistent with the slope and normalization of the \citet{sanders2020} data.
Moreover,  the observed \neiiinwl/\oiinwl \ ratio of the \mstar \ stacks are also consistent with the \citet{sanders2020} $z\simeq3.4$ relation within the uncertainties.
We note that the reason the individual detections lie predominantly above the stacked value is due to selection effects (i.e., the stacks also contain objects that are not individually detected in \neiiinwl \ and/or \oiinwl).
This selection effect is also present in the other two line ratio diagrams, but does not affect the general trends described above.

It can also be seen from Fig. \ref{fig:emission_line_properties} that all of the $z\simeq3.4$ line ratios are elevated with respect to SDSS galaxies at fixed \mstar.
Again, this empirical trend strongly suggests an evolution towards lower \zgas \ at higher redshift, at all stellar masses. 
However, when interpreting the redshift evolution of emission lines, the known evolution towards more extreme \hii \ region conditions at high redshift must also be considered \citep[e.g.,][]{steidel2014, shapley2015}.
Most recent results, as well as the results presented in this paper, suggest that the primary cause of this evolution is the harder ionizing spectra emitted by oxygen-enhanced, low-metallicity (i.e., iron-poor), stars at high redshift \citep[e.g.,][see Section \ref{sec:ofe_results} for further discussion]{steidel2016, strom2017, topping2020b}.
As a consequence, not all of the observed redshift evolution in line ratios can be attributed purely to changes in \zgas \ \citep[e.g.,][]{cullen2016}. 
Nevertheless, these empirical line-ratio diagrams provide useful qualitative indications of the likely evolution of \zgas \ with \mstar \ and redshift, which we discuss quantitatively below.

    \begin{figure}
        \centerline{\includegraphics[width=\columnwidth]{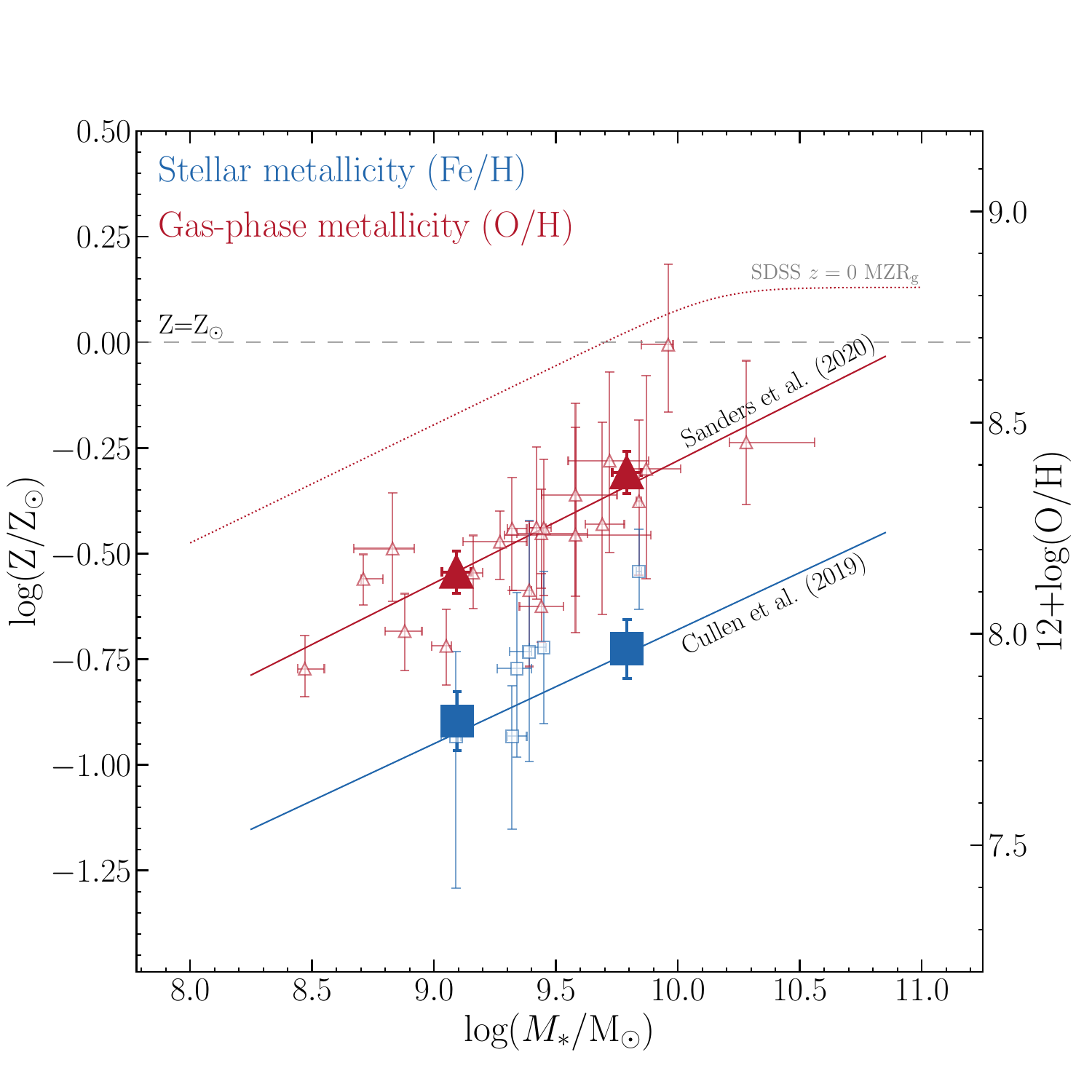}}
        \caption{The mass-metallicity relation for stars (\zstar, blue squares) and nebular gas (\zgas, red triangles) within the NIRVANDELS galaxies at $z\simeq3.4$.
        The small open data points show \zstar \ and \zgas \ for individual galaxies, and the large filled data points show the low-\mstar \ and high-\mstar \ stacks.
        The blue and red solid lines show, respectively, previous determinations of \mzrstar \ and \mzrgas \ at $z\simeq3.4$ from \citet{cullen2019} and \citet{sanders2020}.
        The  \citet{cullen2019} \mzrstar \ has been re-calculated to account for shifts in \mstar \ as described in the text (Section \ref{sec:mzr_gas}).  
        The grey dotted line shows the DIG-corrected \mzrgas \ at $z\simeq0$ from \citet{sanders2020}, and the horizontal dashed grey line indicates the value of solar metallicity on the y-axis.
        On the right-hand side of the y-axis we show the value of log(O/H)+12 for the \zgas \ data (the conversion is given by $12+\mathrm{log(O/H)}=\mathrm{log}(Z_{\star}/\mathrm{Z}_{\odot})+8.69$).
        }
        \label{fig:mzr}
    \end{figure}

\subsubsection{The \mstar-\zgas \ relation}\label{sec:mzr_gas}

The NIRVANDELS $z\simeq3.4$ gas-phase mass-metallicity relationship (\mzrgas) is shown in Fig. \ref{fig:mzr}.
It can be seen that the NIRVANDELS galaxies follow a clear \mzrgas, evident for both the individual galaxies and the stacks.
The difference in log(\zgas) between the high-\mstar \ and low-\mstar \ stacks is $0.25\pm0.08$, representing an increase in \zgas \ of $1.78\pm0.35$ across $\simeq1$ dex in stellar mass ($\simeq 10^9-10^{10}\mathrm{M}_{\odot}$).
It is also clear from Fig. \ref{fig:mzr} that the NIRVANDELS sample is fully consistent with the \citet{sanders2020}  \mzrgas \ determination at $z\simeq3.3$.
This is unsurprising given the similar \mstar-dependence of the emission-line ratios illustrated in Fig. \ref{fig:emission_line_properties}.
Given this excellent agreement, we do not attempt to refit our new data here.
The functional form of the \citet{sanders2020} \mzrgas \ (converted into units of $\mathrm{log}(Z_{\mathrm{g}}/\mathrm{Z}_{\odot})$\footnote{The conversion between $\mathrm{log}(Z_{\mathrm{g}}/\mathrm{Z}_{\odot})$ and the widely-used 12+log(O/H) is simply $12+\mathrm{log(O/H)}=\mathrm{log}(Z_{\rm{g}}/\mathrm{Z}_{\odot})+8.69$.}) is,
\begin{equation}\label{eq:mzr_gas}
\mathrm{log}(Z_{\rm{g}}/\mathrm{Z}_{\odot})=(0.29\pm0.02)m_{10}-(0.28\pm0.03),
\end{equation}
where $m_{10}=\mathrm{log}(M_{\star}/10^{10}\mathrm{M}_{\odot})$.


It can also be seen from Fig. \ref{fig:mzr} that the slope of the $z\simeq3.4$ \mzrgas \ is also fully consistent with the low-mass slope at $z\simeq0$ \citep[$0.28\pm0.01$;][]{sanders2020}.
As a result, there is a constant offset of $\simeq-0.36$ dex in log(\zgas) (a factor 0.44 in \zgas) between $z=0-3.4$ in the relevant mass range ($\simeq10^{8.5}-10^{10.5}\mathrm{M}_{\odot}$).
We note that the size of this offset does critically depend on the \zgas \ calibration used; to this end, the $z\simeq0$ relation of \citet{sanders2020} was derived using a local calibration that should be non-biased with respect to \zgas \ measurements at $z>1$ derived from the \citet{bian2018} calibration, effectively accounting for the harder ionizing radiation field at high redshifts discussed above in Section \ref{subsec:emline_ratios}, and later in Section \ref{sec:ofe_results} (see also \citet{sanders2020} for full details).
Finally we note that, although not shown in Fig. \ref{fig:mzr}, a number of other previous studies at $z>3$ have derived \mstar-\zgas \ relations with a lower overall normalization due to the use of a different metallicity calibration \citep[e.g.,][see \citeauthor{sanders2020} \citeyear{sanders2020} for a full discussion]{maiolino2008,mannuccii2009,troncoso2014, onodera2016}; crucially, however, the slope of these other literature relations are generally consistent with the results presented here, indicating that the scaling of metallicity with stellar mass is independent of the chosen calibration.

\subsection{The stellar mass-metallicity relation at $\mathbf{z\simeq3.4}$}

The NIRVANDELS $z\simeq3.4$ stellar mass-metallicity relationship (\mzrstar) is also shown in Fig. \ref{fig:mzr}.
Again, the individual galaxies and stacks appear to follow a clear \mzrstar. 
The evolution in log(\zstar) between the low-\mstar \ and high-\mstar \ stack is $0.18\pm0.10$ dex (a factor of $1.51\pm0.35$).
In contrast to \mzrgas, the trend for individual objects is less clear due to the fact that the error bars on the individual measurements are large, and the number of objects much reduced.
Nevertheless, these individual measurements are formally consistent with the trend observed in the stacks.

Also shown in Fig. \ref{fig:mzr} is a determination of \mzrstar \ derived from the much larger sample of $\mathrm{N}=681$ VANDELS galaxies presented in \citep{cullen2019}.
The \mstar \ values in \citet{cullen2019} were derived using $H-$ and $K-$ band photometry that had not been corrected for optical emission line contamination, and used a set of SED-fitting assumptions different to the ones that are assumed in this paper.
Therefore, the relation shown in Fig. \ref{fig:mzr} is a re-derivation of the \citet{cullen2019} \mzrstar \ with \mstar \ values derived excluding $H$- and $K$-band photometry and using the same methodology described in Section \ref{subsec:sedfit}.
This relation is in good agreement with our NIRVANDELS data, and is essentially consistent with the original relation shifted to lower \mstar.
The functional form of this \mzrstar \ is given by,
\begin{equation}\label{eq:mzr_star}
\mathrm{log}(Z_{\rm{\star}}/\mathrm{Z}_{\odot})=(0.27\pm0.06)m_{10}-(0.68\pm0.04).
\end{equation}
where $m_{10}=\mathrm{log}(M_{\star}/10^{10}\mathrm{M}_{\odot})$.

There are two striking aspects to the comparison between \mzrstar \ and \mzrgas \ in Fig. \ref{fig:mzr}.
Firstly, \mzrstar \ is offset to lower values at all stellar masses, by $\simeq-0.4$ dex.
This is a result of the fact that \zgas \ traces O/H while \zstar \ traces Fe/H and will be discussed in detail in Section \ref{sec:ofe_results} below.
Secondly, the slopes of both relations are consistent: $\mathrm{d(log}Z)/\mathrm{d(log}M)\simeq0.3$.
Moreover, we note that the low-mass slope of the \mzrstar \ measured in local star-forming galaxies is $\simeq0.38$ \citep{zahid2017}\footnote{This value was calculated by fitting a relation of the form given by equation \ref{eq:mzr_gas} to the \citet{zahid2017} data below $10^{10}$ M$_{\odot}$ (i.e. data below the high mass turnover).}.
This result, combined with the results discussed in Section \ref{sec:mzr_gas}, imply that the power-law slope of the \mstar$-Z$ relation is very similar for both massive stars and \hii \ region gas, and is also not strongly redshift-dependent. 
This in turn would suggest that the physical processes responsible for determining the power-law slope of the \mstar$-Z$ relations below the turnover do not evolve strongly, out to at least $z\simeq4$ (a lookback time of $\simeq12$ Gyr).

\subsection{Enhanced O/Fe ratios at high redshift}\label{sec:ofe_results}

As derived in this paper, \zstar \ traces Fe/H in massive stars while \zgas \ traces O/H in ionized gas (see Section \ref{sec:metallicity_derivation}). 
Crucially, the two quantities are linked via the fact that ionized gas exists in the vicinity of massive stars, and is therefore likely to have very similar chemical abundance properties.
Indeed, due to the short lifetimes of O- and B-type stars, the ionized gas is probably the remnant of their original stellar birth clouds.
In this sense, O/H in ionized gas should also be representative of O/H in massive stars.
Therefore, to a reasonable approximation, combining optically-derived estimates of \zgas \ with FUV-based estimates of \zstar \ traces the O/Fe abundance ratio of the massive stellar populations in star-forming galaxies \citep{steidel2016,topping2020b}.
More generally, because O is an $\alpha-\mathrm{process}$ element, the observed O/Fe enhancement traces the enhancement of $\alpha$ elements, and is also referred to as $\alpha-\mathrm{enhancement}$.

If the NIRVANDELS galaxies have solar-like abundance ratios, we would expect that $Z_{\mathrm{g}}=Z_{\star}$.
However, it is clear from Fig. \ref{fig:mzr} that \zgas \ $>$ \zstar \ across the full range of \mstar.
That is, the O abundances of the NIRVANDELS galaxies are consistently larger than the Fe abundances, implying enhanced O/Fe ratios relative to the solar value.
The values of $\mathrm{log}(Z_{\mathrm{g}}/\mathrm{Z}_{\odot})-\mathrm{log}(Z_{\star}/\mathrm{Z}_{\odot})$ for the low-\mstar \ and high-\mstar \ stacks are $0.35\pm0.09$ and $0.46\pm0.09$, respectively.
These values are both highly significant in themselves ($\approx 4-5 \sigma$), and are consistent with each other within $1.2\sigma$, with an average value of $0.41\pm0.06$. 
Assuming the average value of the two stacks, we find thatthe offset translates into an enhancement of O/Fe relative to the solar value of $\mathrm{(O/Fe)}\simeq 2.57 \pm 0.38 \times \mathrm{(O/Fe)}_{\odot}$.
Within the uncertainties, our results indicate that this offset is constant as a function of \mstar \ between $10^9-10^{10}\mathrm{M}_{\odot}$.

\begin{figure}
        \centerline{\includegraphics[width=\columnwidth]{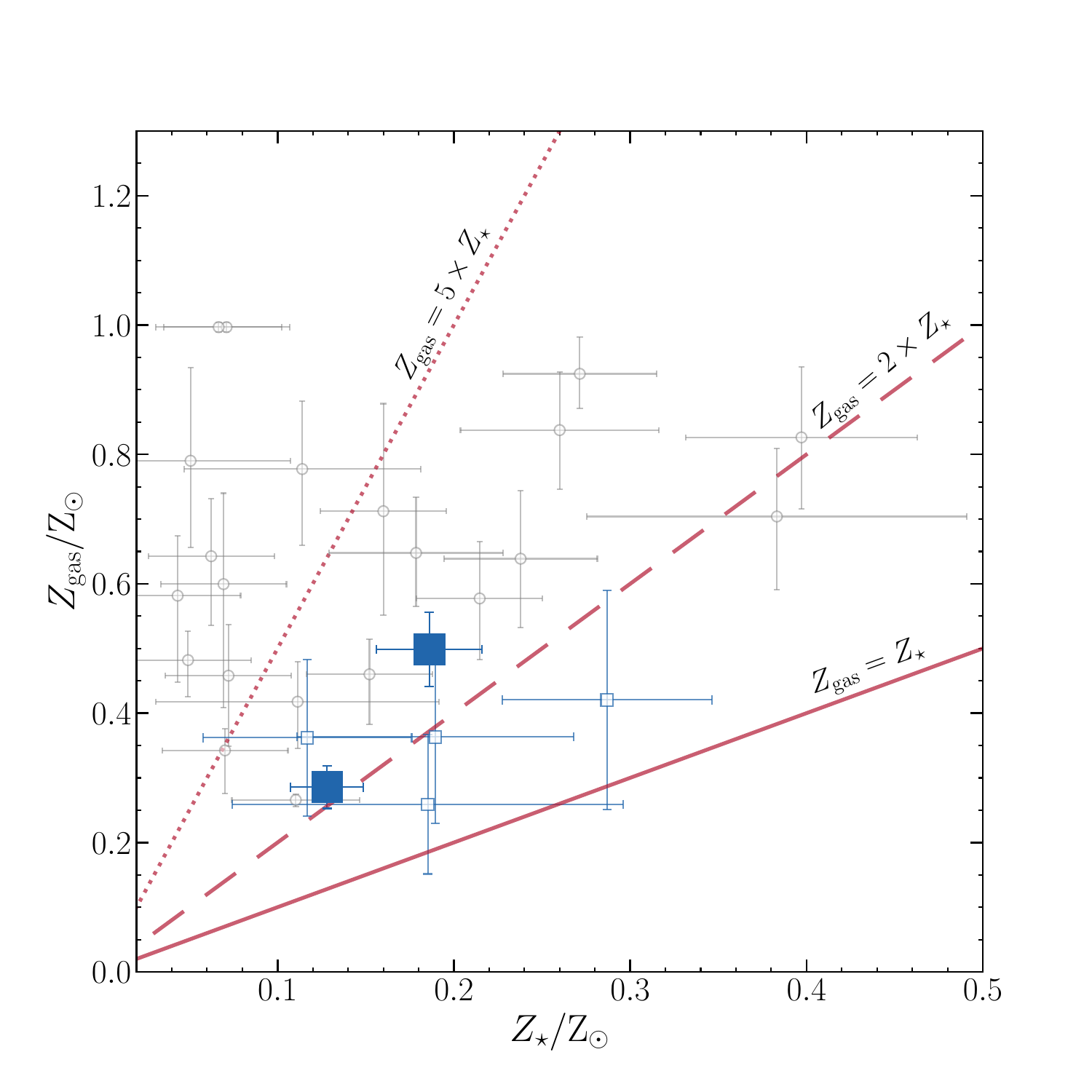}}
        \caption{\zgas \ versus \zstar \ for star-forming galaxies at $z=2-3$.
        Small square data points show the individual galaxies in our sample with a measurement of both quantities.
        Large filled square data points represent the low-\mstar \ and high-\mstar \ stacks.
        The grey circular data points show the data for star-forming galaxies at $z\simeq2.3$ presented in \citet{topping2020b}.
        The various lines show constant \zgas/\zstar \ ratios as indicated by the labels, with the solid line representing the one-to-one relation.}
        \label{fig:zstar_zgas}
    \end{figure}

This result is clarified further in Fig. \ref{fig:zstar_zgas} where we plot, on a linear scale, \zstar \ against \zgas \ (in units of solar metallicity).
It is clear from Fig. \ref{fig:zstar_zgas} that both stacks sit well above the one-to-one relation that would indicate a solar O/Fe ratio (red solid line).
Moreover, each of the four individual galaxies with a measurement of both \zstar \ and \zgas \ also sit above the one-to-one relation. 
Given the large uncertainties the individual results are not highly significant ($<2 \sigma$); nevertheless, based on the more robust stacked measurements, Figs \ref{fig:mzr} and \ref{fig:zstar_zgas} present strong evidence for enhanced O/Fe ratios in typical star-forming galaxies at $z>3$ with $M_{\star} \lesssim 10^{10}\mathrm{M}_{\odot}$.

We note that the evidence presented here for O/Fe enhancement is consistent with other findings in the literature at slightly lower redshift ($z\simeq2.4$, see below), although our results represent the first direct demonstration that (i) O/Fe enhancement exists at $z>3$, and ii) is not strongly dependent on stellar mass below $10^{10}\mathrm{M}_{\odot}$.

\subsubsection{Comparison with the literature}

The most directly comparable works in the literature are the analyses of \citet{steidel2016} and \citet{topping2020a,topping2020b} both of whom performed a similar joint FUV+optical spectral analysis to the one presented in this paper.
\citet{steidel2016} analysed a stack of $30$ star-forming galaxies at $z\simeq2.4$ with \mstar=$10^{9.8}$M$_{\odot}$, finding $\mathrm{(O/Fe)} \simeq 4-5 \times \mathrm{(O/Fe)}_{\odot}$, while \citet{topping2020a,topping2020b} analysed $30$ individual galaxies at $z\simeq2.3$ finding O/Fe enhancements ranging from $1.8 \times \mathrm{(O/Fe)}_{\odot}$ to $> 5 \times \mathrm{(O/Fe)}_{\odot}$.
A direct comparison to the \citet{topping2020b} results is shown in Fig. \ref{fig:zstar_zgas} where it can be seen that the basic result (i.e., $Z_{\mathrm{g}}>Z_{\star}$) is fully consistent with the result presented here.
However, the median enhancement in the \citet{topping2020b} sample ($\approx4.5 \times \mathrm{(O/Fe)}_{\odot}$) is clearly larger than the value we derive ($\approx2.5 \times \mathrm{(O/Fe)}_{\odot}$).
We discuss the main reason for this difference in more detail in the following below.

In addition to the direct method presented here, other indirect methods based on reconciling photoionization models with observed optical emission-line ratios have been used to investigate O/Fe enhancement at high redshift.
In these methods, \zstar \ is constrained mapping Fe/H to the shape of the ionizing continuum spectra used to generate the predicted emission line ratios.
Using such an indirect approach, \citet{strom2018} found $\mathrm{(O/Fe)} \simeq 2.6 \times \mathrm{(O/Fe)}_{\odot}$ for a sample of $150$ main-sequence star-forming galaxies at $z\simeq2.3$.
Similarly, \citet{sanders2020a} also found evidence for significant O/Fe enhancement in a small sample of extreme emission line galaxies with direct temperature-based \zgas \ estimates at $z=1-4$.

We note that although the precise value of the enhancement varies between studies -- due to a combination of selection effects, and systematics related to measuring both \zgas \ and \zstar \ -- the basic result remains consistent. 
That is, star-forming galaxies at high redshift are enhanced in O/Fe, with $\mathrm{(O/Fe)} \gtrsim 2 \times \mathrm{(O/Fe)}_{\odot}$ .
Moreover, as discussed below, there is no known systematic that can easily explain away this result.

\subsubsection{A discussion of systematic effects}\label{subsec:ofe_systematics}

The observed (O/Fe)-enhancement is subject to systematic uncertainties affecting the measurement of both \zgas \ and \zstar.
Sytematics related to the measurement of \zgas \ are well-known and have been discussed extensively in the literature \citep[e.g.][]{kewley2008}.
The \citet{bian2018} calibration used in this paper is arguably the most reliable calibration for galaxies at high redshift, having been built using local analogues of high-redshift galaxies; furthermore, it is known to be consistent with the (currently small) sample of $z\simeq2$ galaxies for which direct-method O/H estimates are available \citep{sanders2020a}.
The \citet{bian2018} calibration returns systematically larger estimates of \zgas \ compared to the commonly-adopted calibration of \citet{maiolino2008}, which has typically been the calibration of choice when determining \zgas \ at $z>3$ \citep{mannuccii2009,troncoso2014, onodera2016}.
However, even assuming the \citet{maiolino2008} calibration, the \zgas \ values reported here only decrease by a factor $\simeq1.4$,  resulting in $\mathrm{(O/Fe)} \simeq 1.8 \times \mathrm{(O/Fe)}_{\odot}$.
We find similar offset using the updated version of these calibrations described in \citet{curti2017}.
In addition, we find that another commonly-used calibration based on the \oiiinwl, \oiinwl \ and \hbeta \ lines - the calibration of \citet{kobulnicky2004} - returns values consistent with \citet{bian2018} for our sample.
In summary, all standard high-redshift \zgas \ calibrations produce estimates of \zgas \ that imply $\mathrm{(O/Fe)} > \mathrm{(O/Fe)}_{\odot}$.

Another crucial systematic related to \zgas \ is the choice of fundamental abundance scale.
The \cite{bian2018} calibration (and indeed the vast majority of empirical \zgas \ calibrations) is tied to temperature-based metallicities measured from auroral lines (often referred to as `direct' method metallicities).
However, there is a well-known discrepancy between \zgas \ determined via the direct method and \zgas \ measured using an alternative method based on oxygen recombination lines (the RL method). 
Estimates of \zgas \ using the RL method are typically $\simeq0.24$ dex higher than \zgas \ estimated using the direct method, with this so-called abundance discrepancy factor (ADF) being roughly constant at all metallicities \citep[][]{peimbert1967,esteban2014}.
It is worth nothing that the results of \citet{steidel2016} discussed above are tied to the RL abundance scale, and this in part explains their larger (O/Fe) estimate. 
Reasons for favoring the RL abundance scale are discussed in detail in \citet{steidel2016} and \citet{strom2017}.
Again, however, accounting for the ADF would only serve to increase \zgas \ (by $\simeq$ factor 2) and therefore strengthen the observed (O/Fe)-enhancement in our sample.

With respect to \zstar, the major source of systematic uncertainty is the choice of SPS model used to fit to the observed stellar continuum.
Along with the SB99 models used in this paper, the Binary Population and Spectra Synthesis (BPASS) models \citep{eldridge2017, stanway2018} are also commonly used in the analysis of high-redshift galaxies.
The primary way in which the BPASS models differ from SB99 is via the inclusion of massive binary star evolution, which can have a strong effect on the predicted UV spectrum, particularly at low metallicities.
In \citet{cullen2019} we found that the BPASSv2.1 models return systematically lower \zstar \ values compared to SB99 by $\simeq 0.1$ dex \citep[see also][]{chisholm2019}.
Fitting our current sample with the latest BPASSv2.2 models (including models at stellar metallicities down to $0.1\%$ solar) yields log(\zstar)$=-3.41\pm0.15$ dex and log(\zstar)$=-2.61\pm0.10$ dex for the low-\mstar \ and high-\mstar \ stacks respectively.
This represents a much larger offset with respect to the SB99 value for the low-\mstar \ stack ($\Delta$ log(\zstar) $\simeq-0.6$ dex) but is consistent with the \zstar \ values reported in \citet{topping2020b} ($\lesssim5\%$ solar).
We conclude that the systematically lower metallicities derived using the BPASS models are the primary cause of the difference between our data and those of \citet{topping2020b} in Fig. \ref{fig:zstar_zgas}.
Crucially, however, use of the BPASS models would only decrease the measured \zstar \ and hence increase the implied (O/Fe)-enhancement of our sample.

Finally, another concern is the depletion of O onto dust grains in the ISM, which would cause \zgas \ to be an underestimate of the true O/H in stars. 
\citet{steidel2016} estimate that O-depletion causes O/H determined from nebular emission lines to be biased by $\simeq -0.09$ dex $(\simeq 20 \%)$ in high-redshift galaxies, with an uncertainty of $\pm 0.05$ dex.
Overall, the effect of O-depletion would be an underestimation of the true gas-phase O/H, resulting in an underestimation of O/Fe.
Again, incorporating the effect of dust depletion would only strengthen the (O/Fe)-enhancement reported here.

In summary, while the exact value of (O/Fe)-enhancement is clearly subject to number of systematic uncertainties, with current estimates in the range $\mathrm{(O/Fe)} \approx 2 - 10 \times \mathrm{(O/Fe)}_{\odot}$, we find no systematic effects that are large enough to explain away the basic result.
Accordingly, we can state with some confidence that the abundance ratios in typical high-redshift star-forming galaxies are $\alpha$-enhanced.


\section{Discussion}\label{sec:discussion}

In this section, we focus on the implications of the main result of this work: O/Fe enhancement (or $\alpha$-enhancement) in high-redshift star-forming galaxies.
As discussed in previous works \citep[e.g.,][]{steidel2016,cullen2019}, O/Fe enhancement is actually expected for galaxies with constant/rising star-formation histories and relatively young ages ($<500$ \Myr \ $-$ 1 \Gyr).
This is a result of the fact that while O is produced predominantly via core-collapse supernovae (CCSNe), Fe is produced via a mixture of \emph{both} CCSNe and Type$-$Ia SNe. 
Therefore, as a result of the delayed onset of Type$-$Ia SNe after the start of star formation, young galaxies will initially be under-abundant in Fe relative to O, only reaching the solar ratio on timescales $\gtrsim 1$\Gyr \ \citep[e.g.,][]{maoz2012,maiolino2019}.
Given the substantial evidence to suggest that star-forming galaxies at $z>2$ are characterised by constant/rising SFHs and ages $\lesssim 1$\Gyr \ \citep{reddy2012,topping2020b}, we therefore expect enhanced O/Fe ratios at this cosmic epoch.
Below we discuss how the observed (O/Fe)-enhancement might relate to local observations, and its consequences with respect to excitation conditions in high-redshift galaxies.

\subsection{Abundance patterns compared to Milky Way stars}\label{subsec:mw_abundance_comp}

It is interesting to consider the link between high-redshift stellar populations and local stellar populations that we know (based on their age) formed at high redshift.
For example, if (O/Fe)-enhancement is ubiquitous among galaxies at $z > 2$, then we would expect to observe enhanced abundance ratios in local stars that formed at these cosmic epochs (i.e., in stars with ages $\gtrsim10$\Gyr).
In the subsequent discussion, we use the following definitions for consistency with local conventions:
\begin{equation}
\mathrm{[Fe/H]} = \mathrm{log(Fe/H)-log(Fe/H)}_{\odot},
\end{equation}
and
\begin{equation}
\mathrm{[O/Fe]} = \mathrm{[O/H]-[Fe/H]}.
\end{equation}
In this work, $\mathrm{log}(Z_{\star}/\mathrm{Z}_{\odot})$ is a proxy for [Fe/H] and $\mathrm{log}(Z_{\mathrm{g}}/Z_{\star})$ is a proxy for [O/Fe].

\begin{figure}
        \centerline{\includegraphics[width=\columnwidth]{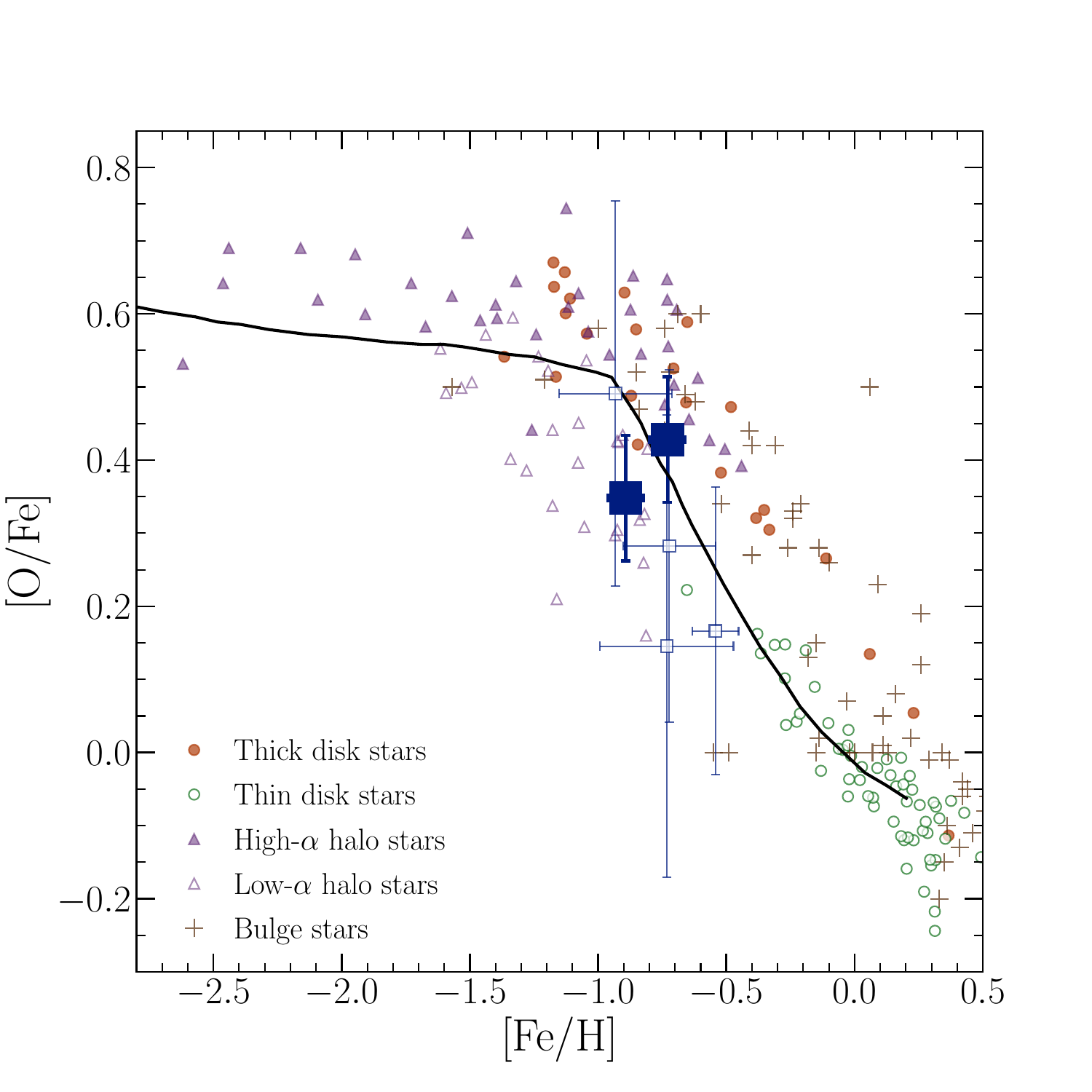}}
        \caption{A comparison between the \oferel \ relation for stars in the Milky Way and star-forming galaxies at $z\simeq3.4$.
        The open square points show the individual galaxies in our sample for which it was possible to measure both \zstar \ and \zgas.
        The filled square points represent low-\mstar \ and high-\mstar  \ stacks (with the low-\mstar \ stack being the most metal-poor i.e., lowest [Fe/H]).
        All other data points represent individual Milky Way stars drawn from the halo, thick disk, thin disk and bulge as indicated in the legend.
        These data have been compiled from \citet{bensby2013} (micro-lensed dwarfs and K-giants), and \citet{zhao2016} and \citet{amarsi2019} (F- and G-type stars).
        The black solid line shows a model for the average \oferel \ relation for stars in the Milky Way, representing the evolutionary history of the galaxy \citep{kobayashi2020}.}
        \label{fig:ofe_feh}
    \end{figure}

Following the argument described above, the relationship between [O/Fe] and [Fe/H] (or more generally any [$\alpha$/Fe]$\mathbf{-}$[Fe/H] relation) can be used to constrain star-formation timescales in galaxies due to the different production timescales for the $\alpha$ elements and Fe.
For example, directly after the onset of star formation, at low [Fe/H], the value of [O/Fe] is set solely by CCSNe yields, as well as the IMF of the first generation of stars \citep{tolstoy2009}.
Current galactic chemical evolution models $-$ constrained by observations of stars in the Milky Way $-$ predict an initial plateau at [O/Fe] $\simeq 0.6$ \citep[Fig. \ref{fig:ofe_feh}; e.g.,][]{kobayashi2020} for stars in our Galaxy.
Other estimates of CCSNe yields give similar values, typically within the range [O/Fe] $\simeq 0.6-0.75$ depending on the assumed stellar metallicity and IMF \citep[e.g.,][]{limongi2003,chieffi2004,nomoto2006}.
Assuming no strong redshift evolution in either the IMF or CCSNe yields as a function of stellar metallicity, this plateau value should be relatively stable at all cosmic epochs.

At later times, Fe enrichment of the ISM becomes enhanced by Type Ia SNe, causing a knee in the relation and subsequent evolution towards [O/Fe] $<0.6$.
In the Milky Way, this knee occurs at [Fe/H] $\simeq -1$.
The sequence in \oferel \ can therefore be thought of in terms of age, with older stars (i.e., those that formed earlier) located at high [O/Fe].
In Fig. \ref{fig:ofe_feh}, we show one recent model of the \oferel \ relation for the Milky Way from \citet{kobayashi2020}, highlighting these trends.

The model of \citet{kobayashi2020}, and other so-called `Galactic Chemical Evolution Models,' are constrained by observations of Milky Way stellar abundance ratios.
For example, studies of FKG stars in the solar neighborhood have revealed a tight correlation between [$\alpha$/Fe] and age, with stars on the knee/plateau having formed $10-12$ Gyr ago \citep[i.e. $z=2-4$;][]{haywood2013,haywood2019}.
These older, high-[$\alpha$/Fe] populations, are predominantly located in the thick disk and halo \citep[][]{bensby2003,haywood2013,zhao2016,kobayashi2020}.
Evidence also exists for a distinct high$-\alpha$ population in the bulge at large scale heights \citep{bensby2013,lian2020}, and is thought to indicate an intimate link between the bulge and disk populations \citep{haywood2018}. 
To illustrate these various populations, we show in Fig. \ref{fig:ofe_feh} the position of a selection of F- and G-type stars drawn from the thin disk, thick disk and halo of the Milky Way using data from \citet{zhao2016} and \citet{amarsi2019}, as well as a selection of micro-{}lensed dwarf and K-giant stars drawn from the bulge taken from \citet{bensby2013}.
As can be seen from Fig. \ref{fig:ofe_feh}, stars in the thick disk and halo are found to have enhanced [O/Fe] while the stars with solar-like abundances are drawn predominantly from the thin disk.
A similar sequence is also evident for the bulge population, with stars at higher [O/Fe] having larger scale heights \citep{amarsi2019}.

     \begin{figure*}
        \centerline{\includegraphics[width=6in]{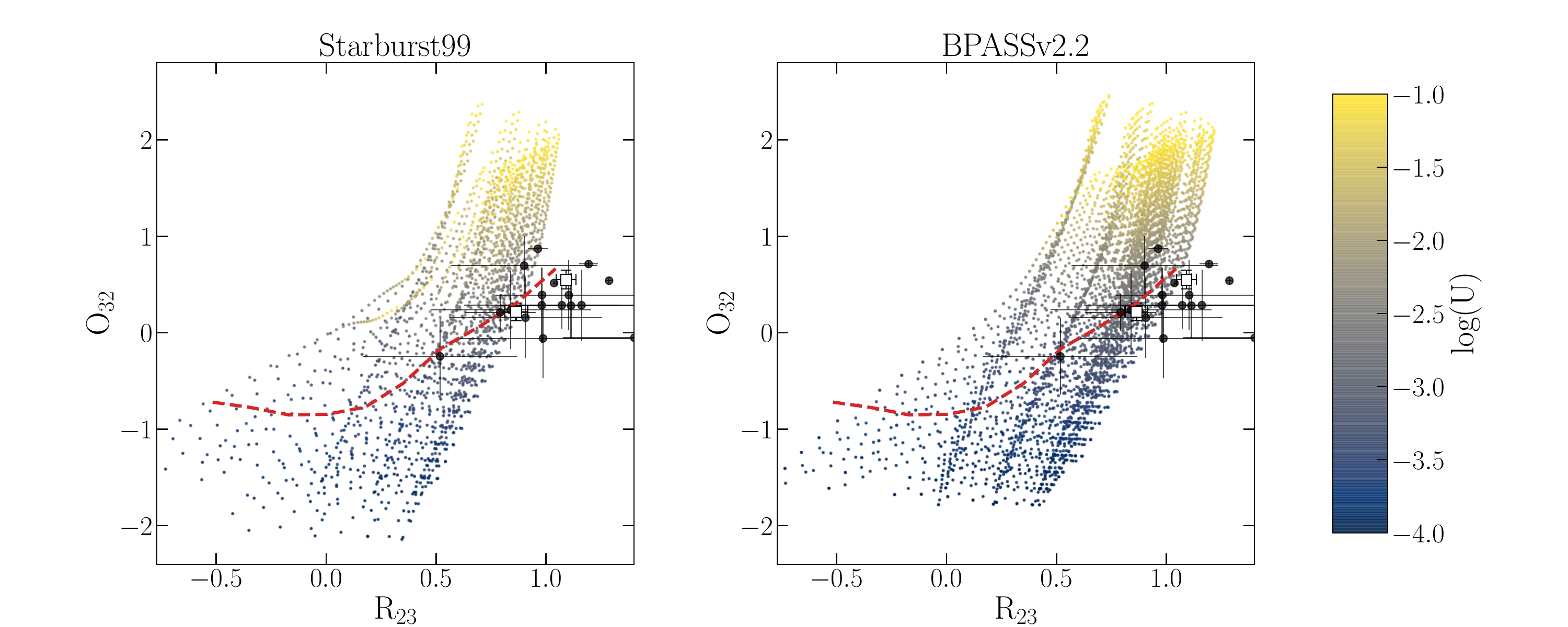}}
        \caption{Comparing theoretical model predictions to observations in the R$_{23}$-O$_{32}$ emission line ratio diagram.
        In each panel the coloured points show the photoionization model grids generated via the method described in Appendix \ref{app:photoion}.
        The left-hand panels show grids generated assuming Starburst99 models for the input stellar ionizing continuum, while the right-hand panels show grids generated assuming the BPASSv2.2 models.
        The colour of the photoionization model points corresponds to the logarithm of the ionization parameter (log($U$)) as shown in the colour bar.
        In each panel the black circular data points show individual galaxies in our sample with detections of the required lines, and the open black squares show the low-\mstar \ and high-\mstar \ stacks.
        The red dashed line shows the running median for a sample of $1050$ local \hii \ regions (see text for description).
        }
        \label{fig:excitation}
    \end{figure*}

It can also be seen from Fig. \ref{fig:ofe_feh} that our observed \oferel \ values are remarkably consistent with the sequence defined locally, corresponding roughly to the location of the knee in the Milky Way sequence dominated by old stars ($\simeq 10-12$ Gyr) drawn primarily from the thick disk, halo and high-$\alpha$ bulge populations of the Milky Way \citep[e.g.][]{haywood2018,haywood2019}.
This similarity suggests that these local $10-12$ Gyr old stellar populations likely formed from similar material (in terms of enrichment properties) to the massive O/B-type stellar populations we observe in the rest-frame FUV spectra of galaxies at $z=3.5$ (lookback time $11.7$ Gyr).
The basic interpretation of this result would be that our $z=3.5$ observations probe the formation of the stellar populations that dominate the largest scale heights (thick disk, subsets of the halo and bulge) in spiral galaxies at $z=0$. 
We note that the potential systematic effects discussed in Section \ref{subsec:ofe_systematics} do not affect this observation.
In fact, converting our \zgas \ measurements to a RL abundance scale would shift [O/Fe] by $\simeq +0.24$ dex, somewhat improving the overlap between our data and the Milky Way halo/bulge population.

These results can also be compared to the element abundance ratios of damped Lyman$-\alpha$ systems (DLAs) at $z>2$, which also exhibit $\alpha$-enhanced abundance ratios at low metallicities ([Fe/H] $\lesssim 1.5$) \citep[e.g.][]{cooke2015,decia2016}.
Interestingly, however, \citet{cooke2015} report a gentle decline towards solar [$\alpha$/Fe] values by [Fe/H] $\simeq -1.0$ for DLAs, in contrast to the significant O/Fe enhancement seen in Fig. \ref{fig:ofe_feh} at similar metallicity. 
Based on their observations, \citet{cooke2015} argue that the typical star-formation and chemical enrichment histories of DLA's are not representative of the galaxy population that formed the Milky Way stellar halo.
Following the same argument,  our results indicate that stellar populations drawn from the main sequence of star-formation at $z\simeq3.4$ are representative of at least a fraction of MW halo stars.

Improved observations of \oferel \ at high redshift have the potential to provide futher constraints on the typical star-formation and chemical evolution histories of early galaxies.
For example, the combined FUV+optical observations presented here can be performed at $z=2-4$, and offer the opportunity to observe the evolution of \oferel \ across $\approx 2$ \Gyr \ of cosmic time (and a wide range in stellar mass), placing unique constraints on the early evolutionary pathways of galaxies and providing unique clues as to the connection between various local and high-redshift stellar populations.

\subsection{Excitation conditions at high redshift}

The observed (O/Fe)-enhancement in $z\simeq3$ NIRVANDELS galaxies also has important implications for our understating of the excitation conditions in high-redshift \hii \ regions.
Below, we briefly examine our results in relation to the much-discussed offset between high-redshift and local star-forming galaxies in classical line ratio diagnostic diagrams.
In addition, we explore whether (O/Fe)-enhancement can resolve the previously-reported difficulty in reproducing $z>3$ emission-line ratios using standard photoionization modelling analyses.

\subsubsection{The evolution of optical emission line ratios with redshift}

It has been known for some time that high-redshift star-forming galaxies are offset from local star-forming galaxies in various nebular emission-line diagnostics diagrams \citep[most notably the N2 BPT diagram; e.g.,][]{steidel2014, shapley2015}.
A number of physical explanations have been invoked to explain this effect, including an evolution in ionization parameter and ISM pressure \citep{cullen2016,kashino2017}, N/O abundance ratios \citep{masters2014,shapley2015,sanders2016} and harder ionizing spectra \citep{steidel2016,strom2017, sanders2020, topping2020b}.
In Fig. \ref{fig:excitation} we show the position of our $z\simeq3.4$ sample in the R$_{23}$ $-$ O$_{32}$ nebular emission line ratio diagram compared to a sample of 1050 local \hii \ regions, where R$_{23}$=log((\oiii+\oii)/\hbeta).
Following \citet{sanders2017} and \citet{jeong2020}, our \hii \ sample is drawn from the catalogue of \citet{pilyugin2016} and supplemented with additional data from \citet{croxall2016} and \citet{toribio2016}.
We choose to use local \hii \ regions (as opposed to SDSS spectra) to avoid the problem of DIG contamination in the integrated spectra of local star-forming galaxies \citep[e.g.][]{sanders2017}.

It can be seen from Fig. \ref{fig:excitation} that, in agreement with much of the previous literature, we find that the $z\simeq3.4$ line ratios are, in general, systematically offset from the median values at $z=0$.
The results presented in this paper, as well as previous observations of (O/Fe)-enhancement at $z>2$, would suggest that one reason for this offset is due to the fact that, at fixed oxygen abundance, star-forming galaxies at $z>3$ have harder ionizing continuum spectra (i.e., lower [Fe/H]) compared to star-forming galaxies in the local Universe.
Indeed, \citet{topping2020a,topping2020b} have shown explicitly how the increase in the excitation conditions within \hii \ regions due to (O/Fe)-enhancement is directly related to the observed N2 BPT offset in $z\simeq2.4$ galaxies.
Our results do not rule out contributions from other physical effects, but highlight the importance of a harder ionizing continuum at high redshift (at fixed O/H).

\subsubsection{Photoionization models}

Another interesting question is whether - under the assumption of $\alpha$-enhanced chemical abundances - current stellar-population synthesis models, combined with photoionization modelling of \hii \ regions, can accurately reproduce the observed $z>3$ nebular emission-line ratios.
Previous literature results have highlighted the fact that photoionization models generally struggle to reproduce the observed (\oiiinwl+\oiinwl)/\hbeta \ ratios in galaxies at $z>3$ \citep[e.g.][]{onodera2016}.
However, these previous comparison have generally been based on the assumption of assumed solar abundance ratios, coupling the metallicity of the ionizing stars to the metallicity of the gas (i.e., \zstar $=$ \zgas, or (O/Fe)=(O/Fe)$_{\odot}$).

In Fig. \ref{fig:excitation}, we also show the results from a photoionization modelling analysis of the R$_{23}$ $-$ \oiiinwl/\oiinwl \ diagram, in which we have relaxed this assumption and decoupled the stellar and gas-phase metallicities.
Full details of the photoionization modelling analysis are given in Appendix \ref{app:photoion}.
It can be seen from Fig. \ref{fig:excitation} that, even when accounting for a harder ionizing continuum at fixed O/H, the photoionization models still fail to reproduce the full range of observed emission-line ratios.
Crucially, the offset between models and observations is evident (although smaller) even when assuming the harder ionizing continuum spectra predicted by the BPASSv2.2 models \footnote{The effects of binary star evolution in the BPASS models results in a significant boost in the ionizing flux at fixed \zstar \ with respect to the Starburst99 models, particularly at \zstar$<0.3$ $\mathrm{Z}_{\odot}$ \ \citep[see e.g.,][]{stanway2016}.}.
These results indicate that, even when accounting for a harder ionizing continuum at fixed O/H (i.e., $\alpha$-enhancement), the shape of the ionizing spectrum predicted by current SPS models is still not sufficient to account for the full range typical emission line ratios observed at $z>3$.

The fact that the BPASSv2.2 models lie closer to the observed line ratios clearly indicates that increasing the strength of the ionizing continuum at fixed \zstar \ helps to alleviate this problem, and that an insufficiently hard ionizing continuum is likely to be contributing to the discrepancy.
Indeed, this conclusion has also frequently been reached with respect to modelling the \heiinwl \ emission line in metal-poor star-forming galaxies across a large range of redshifts \citep[e.g.,][]{plat2019, saxena2020}.
Additionally, the offset may be related to the fact that most photoionization models treat galaxies as single \hii \ regions, when in reality the observed spectra are a line luminosity-weighted averages of many \hii \ regions, each with distinct physical properties.
Our analysis adds to the growing body of literature that suggests that the full range of emission line ratios observed at high-redshift are still not fully captured by simple photoionization models (even when including the effect of $\alpha$-enhanced abundance ratios).
Solving this problem will be crucial for accurately interpreting of the rest-frame optical spectra of galaxies at $z>4$ with \emph{JWST}.
    
\section{Summary and Conclusions}\label{sec:conclusions}

In this paper, we have presented the results of an analysis of the gas-phase and stellar metallicities of $\mathrm{N}=33$ typical star-forming galaxies at $3.0 \leq z \leq 3.8$ drawn from the VANDELS survey.
Using gas-phase metallicity measurements derived from rest-frame optical emission line ratios and stellar metallicity measurements derived from rest-frame FUV continuum fitting, we have investigated the scaling of both parameters with galaxy stellar mass.
In addition, we have directly compared the gas-phase and stellar metallicities of our sample to investigate evidence for (O/Fe)-enhancement in typical star-forming galaxies at $z\simeq3.4$.
Finally, we have discussed how our observations may be related to chemical abundance ratios of stars in the Milky Way.
The main results of this study can be summarized as follows:
\begin{enumerate}

	\item The three rest-frame optical emission-line ratios covered by our MOSFIRE $H$- and $K$-band observations (\oiiinwl/\hbeta, \oiiinwl/\oiinwl, and \neiiinwl/\oiinwl) all show a negative correlation with \mstar.
	The slope of the $z\simeq3.4$ line ratio correlations are consistent with those in the local Universe, but with a higher normalization at fixed \mstar.
	These qualitative trends are to be expected if gas-phase metallicities increase with \mstar \ and decrease with redshift.

	\item Converting the dust-corrected line ratios into gas-phase metallicities (\zgas, or O/H), we observe a clear gas-phase mass-metallicity relationship (\mzrgas) using the calibration of \cite{bian2018}.
	The relation has a gradient of $\mathrm{d(log}Z_{\mathrm{g}})/\mathrm{d(log}M_{\star})\simeq0.3$, with the average value of \zgas \ increasing from $\simeq 0.2$ $\mathrm{Z}_{\odot}$ to $\simeq 0.7$ $\mathrm{Z}_{\odot}$ across the stellar mass range $M_{\star} \simeq 10^{8.5}-10^{10.5}\mathrm{M}_{\odot}$.

	\item We derived stellar metallicities (\zstar, or Fe/H) for our sample based on full FUV continuum fitting in the wavelength range $1250-2000$ {\AA} (rest-frame).
	As with \zgas, we observe an increase in \zstar \ with increasing \mstar.
	The slope of the \mzrstar \ relation is consistent with the \mzrgas \ relation, but \zstar \ is offset to lower values at fixed stellar mass.
	We find \zstar \ increases from $\simeq 0.08$ $\mathrm{Z}_{\odot}$ to $\simeq0.25$ $\mathrm{Z}_{\odot}$ across the stellar-mass range $M_{\star} \simeq 10^{8.5}-10^{10}\mathrm{M}_{\odot}$.

	\item We find an approximately constant offset between \zgas \ and \zstar \ at fixed \mstar, implying enhanced O/Fe relative to the solar value (i.e., $\alpha$-enhancement), with $\mathrm{(O/Fe)} = 2.54 \pm 0.38 \times \mathrm{(O/Fe)}_{\odot}$.
	This result is consistent with the $\alpha$-enhancement observed in galaxy samples at slightly lower redshift \citep[$z\simeq2.4$; e.g.,][]{steidel2016,topping2020b} and is robust against known systematics.
	Within the uncertainties of our observations, there is no strong evidence that (O/Fe)-enhancement is a function of \mstar \ at $M_{\star}<10^{10}\mathrm{M}_{\odot}$.

	\item We compare our measurements to the average \oferel \ sequence of stars in the Milky Way.
	Under the assumption that $\mathrm{log}(Z_{\star}/\mathrm{Z}_{\odot})$ is a proxy for [Fe/H] and $\mathrm{log}(Z_{\mathrm{g}}/Z_{\star})$ is a proxy for [O/Fe], we find the position of our $z\simeq3.4$ sample in the \oferel \ plane is consistent with the knee of the Milky Way sequence.
	In the Milky Way, the knee is predominantly populated by old stars consistent with a formation redshift of $z\sim2-4$ (i.e., $10-12$ \Gyr \ old) found at large scale heights above the plane of the disk.
	Improved constraints on \zstar \ and \zgas \ at high redshift could shed light on the link between high-redshift stellar populations and the structural components in local galaxies using chemical abundance patterns.
	
	\item Finally, we demonstrate that our results are consistent with previous results in the literature suggesting that the observed evolution in rest-frame optical emission-line ratios with redshift is caused, at least in part, by an increase in the hardness of the stellar ionizing continuum at fixed oxygen abundance \citep[e.g.][]{steidel2016, topping2020b}.
	We also show, via a photoionization modelling analysis accounting for (O/Fe)-enhancement, that the full range of observed line ratios at $z>3$ remain difficult to reproduce with standard models.
    Resolving this tension between the models and observations will be crucial for accurately interpreting upcoming spectroscopic data of high-redshift galaxies from \emph{JWST}.

\end{enumerate}

\section*{Acknowledgments}
FC acknowledges the support of the UK Science and Technology Facilities Council.
AC and MT acknowledge the support from grant PRIN MIUR 2017 - 20173ML3WW$\_$001.
RA  acknowledges support from ANID Fondecyt Regular grant 1202007.
We would like to thank Misha Haywood and Annette Ferguson for helpful conversations in relation to Section \ref{subsec:mw_abundance_comp}.
This research made use of \texttt{Astropy}, a community-developed core Python package for Astronomy \citep{astropy2018}, \texttt{NumPy} and \texttt{SciPy} \citep{oliphant2007}, \texttt{Matplotlib} \citep{hunter2007}, \texttt{IPython} \citep{perez2007}, \texttt{SpectRes} \citep{carnall_spectres} and NASA's Astrophysics Data System Bibliographic Services. 

\section*{Data Availability}

The VANDELS spectra used in this work a publicly available and can be accessed via the ESO science portal (\href{http://archive.eso.org/scienceportal/home}{http://archive.eso.org/scienceportal/home}).
All other data contributing to this article will be shared on reasonable request to the corresponding author.



\bibliographystyle{mnras}
\bibliography{nirv} 



\appendix

\section{Photoionization models}\label{app:photoion}

The photoionization model grids discussed in Section \ref{sec:discussion}, and shown in Fig. \ref{fig:excitation}, were generated using Cloudy v17.01 \citep{ferland2017}.
To generate these grids, we assumed a simple plane-parallel geometry and adopted a constant \hii \ region electron density of $n_{e}=350$ cm$^{-3}$ (based on the median \oiinwl \ doublet ratio for our sample and using the conversion given in \citet{sanders2016}). The three variable input parameters were \zgas, \zstar \ and the ionization parameter, $U$.
The ionization parameter sets the intensity of the ionization radiation field and is defined as,
\begin{equation}
U=n_{\mathrm{LyC}}/n_{H},
\end{equation}
where $n_{\mathrm{LyC}}$ is the volume density of hydrogen-ionizing photons and $n_{H}$ is the volume density of hydrogen gas ($n_{H} \approx n_{e}$). 
We varied log($U$) between $-1.0$ to $-4.0$ in steps of $0.1$ dex.
For the gas-phase metallicity, we adopted the following set of values \zgas/Z$_{\odot}=0.05, 0.1, 0.2, 0.3, 0.4, 0.5, 0.6, 0.7, 0.8, 0.9, 1.0,1.25, 1.5$ ($7.4 \leq \mathrm{12+log(O/H)} \leq 8.9$).

In order to account for non-solar O/Fe ratios, we decouple the stellar metallicity (which sets the shape of the ionizing continuum spectrum, and is primarily dependent on Fe/H) from the gas-phase metallicity (tracing O/H).
To generate the input ionizing continuum at a given \zstar \ we adopted two fiducial models, one based on Starburst99 \citep{leitherer2014} and the other BPASSv2.2 \citep{eldridge2017,stanway2018}.
In both cases we assumed continuous star formation with an age of 100 \Myr \ and a \citet{chabrier2003} IMF.
For the Starburst99 models we assumed the same weak-wind Geneva tracks with stellar rotation used to fit the FUV continuum spectra (see Section \ref{sec:metallicity_derivation}). 
The available \zstar \ values for these models are \zstar/Z$_{\odot}=0.07, 0.14, 0.56, 0.99, 2.8$.
For the BPASSv2.2 models we assumed the binary star evolution pathway with an upper-mass IMF cutoff of $300\mathrm{M}_{\odot}$.
The values of \zstar \ available for the BPASSv2.2 models are \zstar/Z$_{\odot}=0.0007, 0.007, 0.07, 0.14, 0.21, 0.28, 0.42, 0.56, 0.70, 0.99, 1.4, 2.1$. 
For each value of \zstar, photoionization grid points were generated across the full range \zgas \ and log($U$) values.
The number of individual grid points for the Starburst99 and BPASSv2.2 grids were therefore 2015 and 4836 respectively.
The resulting photoionization model grids in the R$_{23}$ $-$ O$_{32}$ plane are shown in Fig. \ref{fig:excitation}.

\bigskip

$^{1}$SUPA\thanks{Scottish Universities Physics Alliance}, Institute for Astronomy, University of Edinburgh, Royal Observatory, Edinburgh EH9 3HJ\\
$^2$Department of Physics and Astronomy, University of California, Los Angeles, 430 Portola Plaza, Los Angeles, CA 90095, USA\\
$^3$Department of Physics and Astronomy, University of California, Davis, One Shields Ave, Davis, CA 95616, USA\\
$^4$Department of Physics and Astronomy, University of California, Riverside, 900 University Avenue, Riverside, CA 92521, USA\\
$^5$Instituto de Investigaci\'on Multidisciplinar en Ciencia y Tecnolog\'ia, Universidad de La Serena, Ra\'ul Bitr\'an 1305, La Serena, Chile\\
$^6$Departamento de F\'isica y Astronom\'ia, Universidad de La Serena, Av. Juan Cisternas 1200 Norte, La Serena, Chile\\
$^7$INAF - Osservatorio Astronomico di Bologna, via P. Gobetti 93/3\\
$^8$INAF$-$Osservatorio Astronomico di Roma, Via Frascati 33, I-00040 Monte Porzio Catone (RM), Italy\\
$^9$University of Bologna, Department of Physics and Astronomy (DIFA) 
Via Gobetti 93/2- 40129, Bologna, Italy\\
$^{10}$INAF - Osservatorio Astrofisico di Arcetri, Largo E. Fermi 5, I-50125, Firenze, Italy\\
$^{11}$European Southern Observatory, Karl-Schwarzschild-Str. 2, 86748 Garching b. M\"unchen, Germany\\
$^{12}$INAF-IASF Milano, via Bassini 15, I-20133, Milano, Italy\\
$^{13}$INAF-Astronomical Observatory of Trieste, via G.B. Tiepolo 11, 34143 Trieste, Italy\\
$^{14}$Departamento de Ciencias Fisicas, Universidad Andres Bello, Fernandez Concha 700, Las Condes, Santiago, Chile\\
$^{15}$Space Telescope Science Institute, 3700 San Martin Drive, Baltimore, MD 21218, USA.\\
$^{16}$1Department of Physics and Astronomy, University College London, Gower Street, London WC1E 6BT, UK\\

\bsp	
\label{lastpage}
\end{document}